\documentclass[12pt,preprint]{aastex}

\begin{document}

\title{Fast Molecular Outflows in Luminous Galaxy Mergers:
  Evidence for Quasar Feedback from {\em Herschel}}

\author{S. Veilleux\altaffilmark{1,2,3},
M. Mel\'endez\altaffilmark{1}, 
E. Sturm\altaffilmark{3},
J. Gracia´-Carpio\altaffilmark{3},
J. Fischer\altaffilmark{4},  
E. Gonz\'alez-Alfonso\altaffilmark{5},
A. Contursi\altaffilmark{3}, 
D. Lutz\altaffilmark{3}, 
A. Poglitsch\altaffilmark{3},
R. Davies\altaffilmark{3},
R. Genzel\altaffilmark{3}, 
L. Tacconi\altaffilmark{3}, 
J. A. de Jong\altaffilmark{3},
A. Sternberg\altaffilmark{6}, 
H. Netzer\altaffilmark{6},  
S. Hailey-Dunsheath\altaffilmark{7},  
A. Verma\altaffilmark{8}, 
D. S. N. Rupke\altaffilmark{9},
R. Maiolino\altaffilmark{10, 11},
S. H. Teng\altaffilmark{12},
and E. Polisensky\altaffilmark{4}
}

\altaffiltext{1}{Department of Astronomy, University of Maryland,
  College Park, MD 20742, USA; hkrug@astro.umd.edu,
  veilleux@astro.umd.edu, trippe@astro.umd.edu}

\altaffiltext{2}{Joint Space-Science Institute, University of Maryland,
  College Park, MD 20742, USA}

%\altaffiltext{3}{Astroparticle Physics Laboratory, NASA Goddard Space
%   Flight Center, Greenbelt, MD 20771, USA}

 \altaffiltext{3}{Max-Planck-Institute for Extraterrestrial Physics
   (MPE), Giessenbachstraße 1, 85748 Garching, Germany}

\altaffiltext{4}{Naval Research Laboratory, Remote Sensing Division,
  4555 Overlook Ave SW, Washington, DC 20375, USA}

\altaffiltext{5}{Universidad de Alcal\'a, Departamento de
  F\'{\i}sica y Matem\'aticas, Campus Universitario, E-28871 Alcal\'a
  de Henares, Madrid, Spain}

\altaffiltext{6}{Tel Aviv University, Sackler School of Physics \&
  Astronomy, Ramat Aviv 69978, Israel}

\altaffiltext{7}{Department of Astronomy, California Institute of
  Technology, Pasadena, California, 91125 USA}

\altaffiltext{8}{Department of Astrophysics, Oxford University, Oxford
  OX1 3RH, UK}

\altaffiltext{9}{Department of Physics, Rhodes College, Memphis, TN
  38112, USA}

\altaffiltext{10}{Cavendish Laboratory, University of Cambridge, 19
  J.J. Thomson Ave., Cambridge, CB3 0HE, UK}

\altaffiltext{11}{Kavli Institute for Cosmology, Madingley Road,
  Cambridge, CB3 0HA, UK}

\altaffiltext{12}{Observational Cosmology Laboratory, NASA Goddard Space
  Flight Center, Greenbelt, MD 20771, USA}

\begin{abstract}
  We report the results from a systematic search for molecular (OH 119
  $\mu$m) outflows with {\em Herschel}-PACS\footnote{{\em Herschel} is
    an ESA space observatory with science instruments provided by
    European-led Principal Investigator consortia and with important
    participation from NASA.} in a sample of 43 nearby ($z <$ 0.3)
  galaxy mergers, mostly ultraluminous infrared galaxies (ULIRGs) and
  QSOs.  We find that the character of the OH feature (strength of the
  absorption relative to the emission) correlates with that of the
  9.7-$\mu$m silicate feature, a measure of obscuration in
  ULIRGs. Unambiguous evidence for molecular outflows, based on the
  detection of OH absorption profiles with median velocities more
  blueshifted than $-$50 km s$^{-1}$, is seen in 26 (70\%) of the 37
  OH-detected targets, suggesting a wide-angle
    ($\sim$145$^\circ$) outflow geometry. Conversely, unambiguous
  evidence for molecular inflows, based on the detection of OH
  absorption profiles with median velocities more redshifted than +50
  km s$^{-1}$, is seen in only 4 objects, suggesting a planar or
  filamentary geometry for the inflowing gas.  Terminal outflow
  velocities of $\sim$ $-$1000 km s$^{-1}$ are measured in several
  objects, but median outflow velocities are typically $\sim$ $-$200
  km s$^{-1}$. While the outflow velocities show no statistically
  significant dependence on the star formation rate, they are
  distinctly more blueshifted among systems with large AGN 
    fractions and luminosities [log $(L_{\rm AGN} / L_\odot) \ge 11.8
  \pm 0.3$]. The quasars in these systems play a dominant role in
  driving the molecular outflows. In contrast, the most AGN
  dominated systems, where OH is seen purely in emission, show
  relatively modest OH line widths, despite their large AGN
  luminosities, perhaps indicating that molecular outflows subside
  once the quasar has cleared a path through the obscuring
    material.
\end{abstract}


\section{Introduction}

Gas-rich galaxy merging may trigger major starbursts, lead to the
formation of elliptical galaxies, and account for the growth of
supermassive black holes (BH; e.g., Sanders et al. 1988; Hopkins et
al. 2009).  This merger-driven evolutionary scenario starts with a
completely obscured ultraluminous infrared galaxy (ULIRG). As the
system evolves, the obscuring gas and dust is gradually dispersed,
giving rise to dusty QSOs and finally to completely exposed
QSOs. Powerful winds, driven by the central quasar or the surrounding
starburst, have been invoked to stop the growth of both the BH and
spheroidal component and explain the tight BH-spheroid mass relation
(e.g., Fabian 1999; King 2003; Murray et al. 2005). These winds are
purported to quench star formation in the merger remnants (``negative
mechanical feedback''), creating a population of red gas-poor
ellipticals and explaining the bimodal color distribution observed in
galaxy surveys (e.g., Kauffmann et al. 2003). There is growing
observational support for these winds: e.g., most galaxies with high
star formation rate (SFR) densities show signatures of outflows, both
locally and at high redshifts (e.g., Heckman 2002; Veilleux, Cecil, \&
Bland-Hawthorn 2005; Chen et al. 2010; Weiner et al. 2009; Steidel et
al. 2010; Kornei et al. 2012; Martin et al. 2012; Newman et al. 2012
and references therein). The local outflows are often spatially
resolved, allowing determination of the mass outflow rates ($\sim$
0.1-5 $\times$ SFR) and kinetic energies ($\sim$10$^{56 - 58}$ ergs in
ULIRGs; Rupke, Veilleux, \& Sanders 2002, 2005a, 2005b, 2005c; Martin
2005, 2006). These winds are primarily driven by the starburst rather
than the active galactic nucleus (AGN), except in late-stage mergers
with quasar signatures where velocities $\ga$1000 km~s$^{-1}$ (cf.\
100-400 km~s$^{-1}$ in other systems) are sometimes observed (e.g.,
Rupke et al. 2005c; Krug, Rupke, \& Veilleux 2010; Krug et al. 2013,
in prep.).

Studies conducted in the past three years are shedding new light on
the wind phenomenon in the local universe: (1) Early results from our
{\em Herschel} guaranteed time key program SHINING (PI Sturm) have
revealed far-infrared (FIR) OH features with P-Cygni profiles
indicative of massive molecular outflows in a number of ULIRGs
(Fischer et al. 2010; Sturm et al. 2011, hereafter F10 and S11,
respectively; also Gonzalez-Alfonso et al. 2012, 2013). In a few of
these objects, particularly those with dominant AGN such as Mrk~231,
the nearest quasar known, the terminal outflow velocities exceed 1000
km s$^{-1}$ and the outflow rates (up to $\sim$1000 $M_\odot$
yr$^{-1}$) are several times larger than the infrared-based SFRs.  (2)
Independent, spatially resolved CO-emission observations of Mrk~231
with the IRAM/PdB mm-wave interferometer (Feruglio et al. 2010; Cicone
et al. 2012; it has also been mapped in HCN, HCO$^+$, and HNC by Aalto
et al. 2012) have confirmed this outflow and deduced mass outflow
rates of $\sim$700 M$_\odot$ yr$^{-1}$, far larger than the on-going
infrared-based SFR ($\sim$160 M$_\odot$ yr$^{-1}$) in the host
galaxy. Further mm CO-observations have now been done by our group in
a number of the objects discussed here, finding similar CO outflows
(Cicone et al. 2013). Remarkably, the CO outflow in Mrk~231 partially
overlaps spatially and kinematically with blueshifted optical Na~I~D
5890, 5896 \AA\ absorption features detected out to $\sim$2-3 kpc from
the nucleus (Rupke et al. 2005c).  (3) Our high-resolution Gemini/IFU
observations have revealed that the Na I~D outflow in Mrk~231 is
wide-angle, thus driven by a QSO wind rather than a jet (Rupke \&
Veilleux 2011, hereafter RV11).  Similar wide-angle high-velocity
outflows have now been mapped at high resolution in a number of other
local ULIRGs (Rupke \& Veilleux 2013a, hereafter RV13, and 
  2013b).

These powerful outflows may be the long-sought ``smoking gun" of
quasar mechanical feedback that clears out the molecular disk formed
from dissipative collapse during the merger. The tentative trend of
increasing OH terminal outflow velocity with increasing AGN luminosity
identified by S11 adds support to this idea, but it is based on only 6
ULIRGs (and NGC~253, a regular starburst galaxy). To properly test
this idea, one needs to examine a statistically representative sample
of ULIRGs and study the properties of their molecular outflows
uniformly and systematically. In this paper, we report the first
results from such a study, focusing on the directly measured outflow
velocities [in a later paper, we will report on our analysis of the
model-dependent dynamical quantities (masses, momenta, and energies)
in the subset of objects for which we have multiple OH
transitions]. The sample used for this analysis is described in \S
2. The observations and methods used to reduce and analyze the data
are discussed in \S 3. The results from this analysis are presented in
\S 4 and their implications are discussed in \S 5. The main
conclusions are summarized in \S 6. Throughout this paper, we adopt
$H_0$ = 70 km s$^{-1}$ Mpc$^{-1}$, $\Omega_M$ = 0.3, and
$\Omega_\Lambda$ = 0.7. We also adopt the standard convention that
approaching material has a negative velocity with respect to the
systemic velocity of the host galaxy due to Doppler shift.

\section{Sample}

\subsection{Sample Selection}

As mentioned in \S 1, the first results from the SHINING GTO survey of
ULIRGs (S11) revealed a tentative trend between OH outflow velocities
and AGN luminosities, but this was largely driven by the velocities
measured in Mrk~231 and F08572+3915, the only two warm (large
$f_{25}/f_{60}$ flux ratio), quasar-dominated, late-stage mergers in
the S11 sample. It quickly became clear that the rest of the SHINING
GTO data would not be able to address this question adequately because
the GTO sample largely probes the early stages of the merger after the
first peri-passage: 20 of the 23 (87\%) GTO ULIRGs are cold (small
$f_{25}/f_{60}$ ratio) pre-mergers or recent mergers, where the
bolometric and mechanical luminosities are still dominated by the
nuclear starburst (Veilleux et al. 2009b, hereafter V09).  A cycle 1
open-time program (OT1$\_$sveilleu$\_$1, PI Veilleux)
% proposal was granted 32.5 hrs 
has allowed us to expand the original GTO sample to include a set of
warm quasar-dominated ULIRGs. These objects were selected using four
criteria: 1. The targets had to be part of the {\em QUEST} sample of
local ($z < 0.3$) luminous mergers: either ULIRGs or QSOs (Veilleux
2012 and references therein). 2. Their bolometric luminosity had to be
dominated by the quasar based on the {\em Spitzer} data or,
equivalently, {\em IRAS}-band 25-to-60 $\mu$m flux ratio $f_{25}/f_{60}
\ga 0.15$ (this criterion also automatically selects late-stage, fully
coalesced mergers; see V09).  3. High signal-to-noise ratio (S/N
$\sim$30-40) Na~ID absorption spectra had to be planned or available
for each object. This criterion did not bias the sample in any way
since these objects were selected randomly based on visibility at the
time of the ground-based observations. 4. They had to have $f_{100} >
1$ Jy so that high S/N in the continuum could be reached in a
reasonable amount of time with {\em Herschel}-PACS. These criteria
resulted in a sample of 18 objects, 3 of which were already part of
the SHINING GTO survey of ULIRGs.  This sample size was a good match
to the existing 20 cold pre-merger ULIRGs in SHINING.

Since all ULIRGs in the SHINING survey and its OT1 extension were
selected to have {\em IRAS}-band 100 $\mu$m fluxes $f_{100} >$ 1 Jy to
more easily probe OH 119 $\mu$m, this sample did not contain any
``classic'' IR-faint QSOs in the critical late merger phases when the
quasar has finally gotten rid of its natal ``cocoon'' and the effects of
feedback are predicted to subside (Narayanan et al. 2008; Hopkins et
al. 2009). So we put in a successful open-time request
% for 37.2 hrs
in Cycle 2 (OT2$\_$sveilleu$\_$4, PI Veilleux) to further expand the
sample by including objects meeting the following three additional
criteria: 1. IR-to-bolometric luminosity ratios smaller than 0.6 {\em
  i.e.}  IR-fainter than all ULIRGs in the SHINING and OT1 programs
(IR here and throughout this paper refers to 8 -- 1000 $\mu$m),
2. High-S/N ($\sim$30-40) Na~ID and FUV absorption spectra had to be
planned or available for each object.
% Again, the Na~ID criterion did
% not bias the sample in any way since these objects were selected
% randomly based on visibility at the time of the ground-based
% observations. 
The new FUV criterion favored the ``classic'' UV-bright QSOs. 3. The
$f_{100}$ threshold was lowered by a factor of 2 down to $>$ 0.5 Jy to
accommodate these IR-fainter systems. This factor of 2 was chosen to
provide enough objects in the OT2 sample without requiring unrealistically
long exposure times per target.
% , while still requiring a manageable amount of time with {\em
%   Herschel}-PACS ($\le$14 hrs). 
These criteria resulted in a sample of 5 objects.  While this sample
size is small, recall that the main objective of this OT2 extension is
to anchor the merger sequence at the latest (IR-faintest) stages that
could be probed with {\em Herschel}. This is a merger phase that has
not been explored with {\em Herschel} until now so each of these
objects provides important new information.

In summary, the sample consists of 43 objects, 23 from the GTO sample,
15 from the OT1 program, and 5 from the OT2 program.  All of these
objects show signs of on-going or recent interactions (NGC~4418 =
F12243$-$0036, a GTO target, is not a major merger but it is infrared
luminous so we include it in the present sample).  We include here the
ULIRGs from S11 again, applying the same fitting method as for the
rest of the sample, in order to be complete and consistent. The main
objective of the present study is to search for trends with AGN and
host properties as well as age across the merger sequence (after the
first encounter, post-LIRG phase).  In \S 3, we will subdivide these
43 mergers into various subcategories (e.g., morphology, starburst
power, AGN luminosity) to try to find the driving parameter of the
wind properties. Our {\em Spitzer} results (V09) have shown that black
hole accretion, in addition to depending on the merger phase, also has
a strong chaotic/random component, so a sample size substantially
smaller than 43 objects would make it hard to draw statistically
significant conclusions.

\subsection{Properties of the Sample Galaxies}

The properties of our sample galaxies are listed in Table 1.  The
notes to Table 1 briefly explain the meaning of each of these
quantities. The luminosities listed in Table 1 rely on a number of
assumptions that deserve further explanation.  Following
V09, the bolometric luminosities for ULIRGs were estimated to be
$L_{\rm BOL}$ = 1.15 $L_{IR}$, where $L_{\rm IR}$ is the infrared
luminosity over 8 -- 1000 $\mu$m (Sanders \& Mirabel 1996), and
$L_{\rm BOL} = 7 L({\rm 5100~\AA}) + L_{\rm IR}$ for the PG~QSOs
(Netzer et al. 2007). The starburst and AGN luminosities were next
calculated from
\begin{eqnarray}
L_{\rm BOL} & = & L_{\rm AGN} + L_{\rm SB} \\
           & = & \alpha_{\rm AGN}~L_{\rm  BOL} + L_{\rm SB},
\end{eqnarray}
where $\alpha_{\rm AGN}$ is the fractional contribution of the AGN to
the bolometric luminosity, hereafter called the ``AGN fraction'' for
short. In V09, we measured $\alpha_{\rm AGN}$ in the {\em QUEST}
ULIRGs and QSOs using six independent methods that span a range in
wavelength and give consistent results within $\sim$ $\pm$10\%--15\%
on average. These six methods relied on (1) the [O~IV]/[Ne~II] ratio,
(2) the [Ne~V]/[Ne~II] ratio, (3) the PAH 7.7 $\mu$m equivalent width,
(4) the modified version of the Laurent et al. (2000) diagram which
plots the 14-16 $\mu$m to 5.3-5.8 $\mu$m continuum ratio versus the
PAH 6.2 $\mu$m equivalent width, (5) the PAH-free, silicate-free 5 --
25 $\mu$m to FIR continuum ratio, and (6) the 15-to-30 $\mu$m
continuum ratio, $f_{\rm 15}/f_{\rm 30}$.

Unfortunately, several objects in the SHINING GTO sample are not part
of the {\em QUEST} sample and therefore do not have such accurately
measured AGN contribution to the bolometric luminosity. In the
following discussion, we rely solely on the rest-frame $f_{\rm
  15}/f_{\rm 30}$ continuum ratio, which is available for all the
objects in the {\em Herschel} sample, to quantify the AGN
contributions to the bolometric luminosities of our systems. This
ratio was found by V09 to be more tightly correlated with the
PAH-free, silicate-free MIR/FIR ratio and the AGN contribution to the
bolometric luminosity than any other {\em Spitzer}-derived continuum
ratio.

Following V09, we use $f_{30}/f_{15}$ as a surrogate of the PAH-free,
silicate-free MIR/FIR ratio and adopt log($f_{30}/f_{15}$) = 0.20 and
1.35 as the zero points for the ``pure'' AGN and the ``pure''
starburst ULIRG, respectively (note that V09 make a distinction
between ``pure'' starburst ULIRGs and ``normal'' star-forming
galaxies). The AGN contribution is calculated from a linear
interpolation between these two extremes. The zero point for the pure
AGN corresponds to the average $f_{30}/f_{15}$ ratio of FIR-undetected
PG QSOs, while the zero point for the pure starbursts is calculated
from the ten ULIRGs with the largest $f_{30}/f_{15}$ ratios. We used
the correction factors listed in Table 10 of V09 to transform 15
$\mu$m luminosities into bolometric luminosities: log[$L_\nu$(15
$\mu$m)/$L_{\rm BOL}$] = --14.33 for a pure AGN and --14.56 for a pure
starburst ULIRG. The uncertainty on $\alpha_{\rm AGN}$ is estimated to
be $\pm$20\% on average, but is likely higher for strongly buried
sources with significant FIR extinctions (e.g.,
% NGC~4418 = F12243$-$0036, 
Arp~220 = F15327$+$2340). We return to the issue of FIR extinction in
\S 4.6 and \S 5.

Figure 1 shows the distributions of redshifts, starburst luminosities,
AGN fractions, and AGN luminosities for all 43 objects in our sample.
This figure emphasizes the broad range of properties of our sample
galaxies.  The starburst and AGN luminosities are independent of each
other {\em i.e.} we find no correlation or anti-correlation between
the starburst and AGN luminosities in our sample of galaxies, but
perhaps not surprisingly objects with the largest starburst 
  (AGN) luminosities ($\ga$ 10$^{12}$ $L_\odot$) tend to have lower
(higher) AGN fractions.

\section{Observations, Data Reduction, and Spectral Analysis}

\subsection{Observations}

All of the data in this paper were obtained with the PACS far-infrared
spectrometer (Poglitsch et al. 2010) on board {\em Herschel} (Pilbratt
et al. 2010).  Several systems in the OT1 and OT2 sub-samples are
relatively faint in the FIR so an exhaustive multi-line PACS survey
was not possible.  Contrary to the GTO data, which generally cover the
ground-state OH 119 $\mu$m $^2\Pi_{3/2}$ $J$ = 5/2 -- 3/2 rotational
$\Lambda$-doublet transitions, the high-lying ($E_{\rm lower}$ = 290
K) 65 $\mu$m $^2\Pi_{3/2}$ $J$ = 7/2 -- 9/2 and ($E_{\rm lower}$ = 120
K) 84 $\mu$m $^2\Pi_{3/2}$ $J$ = 7/2 -- 5/2 rotational
$\Lambda$-doublet transitions, and the ground-state cross-ladder 79
$\mu$m $^2\Pi_{1/2}$ -- $^2\Pi_{3/2}$ $J$ = 1/2 -- 3/2 rotational
$\Lambda$-doublet transitions, the OT1 and OT2 data focus on a single
OH feature, the OH 119 $\mu$m doublet (and $^{18}$OH 120 $\mu$m
counterparts). The value of this feature as a wind diagnostic was
beautifully demonstrated in F10 and S11 and our early GTO data. It is
the strongest OH transition in Mrk~231 and most of the GTO targets. It
is positioned in wavelength near the peak spectroscopic sensitivity of
PACS. Finally, it is seen in absorption more often than the 79 $\mu$m
feature and thus is more likely to provide an unambiguous signature of
gas motion (rotation, inflow, or outflow).

PACS was used in range scan spectroscopy mode in high sampling
centered on the redshifted OH 119 $\mu$m + $^{18}$OH 120 $\mu$m
complex with a velocity range of $\sim$8000 km s$^{-1}$ (rest-frame
118-121 $\mu$m) to provide enough coverage on both sides of the OH
complex for reliable continuum placement. The resulting PACS spectral
resolution is $\sim$270 km s$^{-1}$. The total amount of time spent to
carry out these observations (including overheads) was 13.0, 32.5,
and 37.2 hrs during GT, OT1, and OT2, respectively. The on-target
exposure times were adjusted according to the continuum level under
OH, estimated from an extrapolation of the {\em Spitzer} mid-infrared
+ {\em IRAS} FIR energy distribution.  A small chopper throw of
1.5$\arcmin$ was used in most cases.
% Standard chopping/nodding mode with 
The observing time for each target, including all overheads, is listed
in the last column of Table 1.

\subsection{Data Reduction}

All of the PACS data were reduced in the same way as in S11.  As
described there, the data reduction was done using the standard PACS
reduction and calibration pipeline (ipipe) included in HIPE
6.0. 
% 6.0.2055
The spectra were normalized to the telescope flux (which dominates the
total signal) and re-calibrated it with a reference telescope spectrum
obtained from dedicated Neptune observations during the Herschel
performance verification phase. All of our objects are point sources
for PACS. In the following, we use the spectrum of the central
9$^{\prime\prime}$ $\times$ 9$^{\prime\prime}$ spatial pixel (spaxel)
only, applying the point-source correction factors (PSF losses) as
given in the PACS documentation. We have verified this approach by
comparing the resulting continuum flux density level to the continuum
level of all 25 spaxels combined (which is free of PSF losses and
pointing uncertainties). In all cases the agreement is excellent,
however the central spaxel alone provides better S/N.

The reduced spectra were next smoothed using a Gaussian kernel of
width 0.05 $\mu$m ({\em i.e.} about half a resolution element) to
reduce the noise in the data before the spectral analysis. A spline
was fit to the continuum and subtracted from the spectra; these
continuum-subtracted spectra were used for the subsequent spectral
fitting.

\subsection{Spectral Analysis}

The profiles of the OH 119.233, 119.441 $\mu$m doublet were modeled
using four Gaussian components. The fits were carried out using {\em
  PySpecKit}, an open-source, extensible spectroscopic analysis
toolkit for astronomy (Ginsburg \& Mirocha 2011). This tool uses the
Levenberg-Marquardt technique to solve the least-squares problem in
order to find the best fit for the observations.\footnote{By default
  PySpecKit implements the Levenberg-Marquardt algorithm via MPFIT
  (Markwardt, C. B. 2009). } Each line of the OH doublet was fitted
with two Gaussian components characterized by their amplitude (either
negative or positive), peak position, and standard deviation (or,
equivalently, FWHM). The separation between the two lines of the
doublet was set to 0.208$\mu$m in the rest-frame ($\sim$520 km
s$^{-1}$) and the amplitude and standard deviation were fixed to be
the same for each component in the doublet. Overlap effects between
the various components were ignored for simplicity (e.g., the emission
component in P Cygni profiles is not affected by the absorption
component in the foreground. This is discussed further in the next
paragraph). The $^{18}$OH 120 $\mu$m doublet is detected in a number
of objects but is too far to the red to significantly affect the
profiles of the OH 119 $\mu$m feature. Likewise, contamination from
$^{17}$OH, H$_2$O$^+$, CH, CH$^+$, and CO (22 -- 21) is considered
unlikely since these lines are undetected even in the high-S/N
spectrum of Mrk~231 (see Figure 2a of Gonzalez-Alfonso et al. 2013).

We find that three distinct scenarios apply to our data: 1) Pure OH
absorption, 2) pure OH emission, and 3) P~Cygni profiles (our data
reveal no new evidence for inverted P~Cygni OH profiles; this
interesting result is discussed further in \S 4.3). In scenario \#1,
there is no evidence for any OH emission and each line of the OH
doublet is fitted with two absorption components, tracing both the
stronger low-velocity component of the outflow and the fainter
high-velocity component (e.g., Mrk~231; Gonzalez-Alfonso et al. 2013).
Scenario \#2 is treated similarly. In this scenario, there is no
evidence for any OH absorption feature and two Gaussian components are
used to model each of the line in the OH doublet.  Scenario \#3 is
more subtle. We first tried to fit the P~Cygni profiles with 3 or 4
components per line (1-2 blueshifted absorption components and 1-2
redshifted emission components), but we quickly realized that these
fits were underconstrained. We had to settle on fitting these profiles
with a single blueshifted absorption component and a single redshifted
emission component {\em i.e.} no attempt is made to capture the
fainter high-velocity component of the outflow. This has the
  potential to underestimate the outflow velocities derived from this
method (making the velocities derived from the absorption feature more
positive than they should be). There is also the risk of a
degeneracy between the emission and absorption components since in
principle they each could be unrealistically strong as long as they
cancel each other to reproduce the spectrum. In the radiative transfer
models of Gonzalez-Alfonso et al. (2012, 2013), overlap effects
between the background emission component and the foreground
absorption component prevent this degeneracy, virtually zeroing out
the emission component(s) over the range of velocities of the
absorption component(s).  To approximate these overlap effects, the
fit in some cases is further constrained by requiring the
absorption feature to not be deeper than the observed absorption.
This assumption was made for six objects (F09022$-$3615,
  F10565$+$2448, F12072$-$0444, F12112$+$0305, F14348$-$1447, and
  F23389$+$0300), but in the end there are only two cases where the
  emission component affects the absorption from the blue (119.233 um)
  line of the doublet: F12112$+$0305 and F23389$+$0300.  However, even
  in these objects, the emission contribution is weak at the position
  of this line and does not affect the measured velocities by more
  than the typical measurement uncertainties ($\pm$50 km~sec$^{-1}$;
  see below), based on comparisons with unconstrained fits.  In \S
  4.3, we compare the velocities measured in scenario \#2 with those
  measured in scenario \#1 and find no significant shift, suggesting
  that the assumptions made in scenario \#2 have no systematic effect
  on the velocity measurements.  Nevertheless, we use caution and
  distinguish the absorption velocities derived under scenario \#2
  from those measured in scenarios \#1 and \#3 in many of the figures
  presented in \S 4.

These fits were first used to quantify the strength and nature
(absorption {\em vs} emission) of this feature: (1) the total flux and
equivalent width of the OH 119.441 $\mu$m line, adding up all of the
absorption and emission components, (2) the flux and equivalent width
of the absorption component(s) used to fit this line, and (3) the flux
and equivalent width of the emission component(s) used to fit this
line. 

We next characterized the OH profile by measuring a few characteristic
velocities from our fits: (1) $v_{50}$(abs), the median velocity of
the fitted absorption profile {\em i.e.} 50\% of the absorption takes
place at velocities above (more positive than) $v_{50}$(abs), (2)
$v_{84}$(abs), the velocity above which 84\% of the absorption takes
place, (3) $v_{50}$(emi), the median velocity of the fitted emission
profile {\em i.e.} 50\% of the emission takes place at velocities
below (less positive than) $v_{50}$(emi), and (4) $v_{84}$(emi), the
velocity below which 84\% of the emission takes place. We did not
extend our fit-based analysis beyond 84\% of the absorption / emission
profiles because the wings of the OH profiles may not be well captured
by our simple two-component fits [especially in the case of P Cygni
profiles where a single Gaussian was used to model the absorption (or
emission) feature of each line in the doublet]. For comparison with
S11, we also estimated the terminal outflow velocity, $v_{\rm max}$,
from the maximum extent of the blueshifted wing of the OH 119.233
$\mu$m absorption profile (it was not measured in the case of pure OH
emission profiles). This quantity was measured independently by
three members of our team to try to reduce possible biases, but it is
admittedly more subjective and uncertain than the other velocity
measurements, as noted below.

There are a number of potential sources of uncertainties when
measuring the equivalent widths, fluxes, and velocities from our
data. The wavelenth uncertainty depends on mispointing. Any
mispointing may also have an effect on the line profiles. We checked
for this and found the effect not to apply in our cases. We find
instead that the uncertainties on these measurements are dominated by
the placement of the continuum. To estimate this source of
uncertainty, we re-measure these quantities assuming different
continuum shapes (spline {\em vs} polynomial) and positions.  We find
that the typical uncertainties on the equivalent widths and fluxes are
$\pm$20\% and $\pm$50 km s$^{-1}$ on the velocities, except for
$v_{\rm max}$ which is much more uncertain ($\pm$200 km s$^{-1}$).
{\em For this reason, we give considerably more weight to results
  based on $v_{50}$ and $v_{84}$ in the following discussion.}

\section{Results}

Figure 2 shows the fits to the OH 119 $\mu$m profiles. The OH 119
$\mu$m parameters derived from these fits are tabulated in Table 2
along with their uncertainties. The average and median values of
  these parameters are listed at the end of this table.
The meaning of each parameter is discussed in \S 3.3 and the notes to
Table 2.  Note that the fluxes and equivalent widths in that table
need to be multiplied by a factor of 2 when considering both lines of
the doublet. In this section we compare these results with the other
galaxy properties tabulated in Table 1.

\subsection{Detection Rate of the OH 119 $\mu$m Feature}

The 119 $\mu$m OH doublet was detected in 37 of the 43 objects in our
sample (86\%). The OH detection rate shows no obvious dependence on
the properties of the objects (Figure 3). A closer inspection of the
data indicates that the six objects without OH detection have
systematically lower S/N ratios ($\la$ 20 -- 30) in the FIR continuum,
and therefore less stringent limits on the OH 119 $\mu$m equivalent
widths, than the others. Two of them are IR-faint QSOs from the OT2
sample (PG~1126$-$041
% PG~1351$+$640, 
and PG~2130$+$099). The other four are part of our OT1 sample of warm
ULIRGs (F13305$-$1739, F15206$+$3342,
% F22491$-$1808, 
F23128$-$5919, and F12265$+$0219 = 3C 273). These six objects are
excluded from the rest of the analysis.

\subsection{Nature of the OH 119 $\mu$m and Mid-Infrared Silicate
  Features}

Of the 37 OH 119 $\mu$m detections, 17 are seen purely in
absorption\footnote{Some of these 17 objects (e.g., F05024$-$1941,
  F15462$-$0450, and F20551$-$4250) may have faint redshifted emission
  but our simple fitting procedure was not successful in capturing
  this emission.}, 15 show absorption + emission composite
profiles, and 5 are purely in emission. Figure 4 shows no obvious
trend between the total equivalent width of the OH feature and AGN
fraction, starburst luminosity or AGN luminosity, except for the fact
that all four objects with AGN fractions $\alpha_{\rm AGN}$
  $\ga$90\% show pure OH emission. This result is consistent with the
detection of OH 119, 79, and 163 $\mu$m emission in the prototypical
Seyfert 2 galaxy NGC~1068 (Spinoglio et al. 2005; Hailey-Dunsheath et
al. 2012). Preliminary results on the {\em Swift} BAT AGN sample
(OT2$\_$sveilleu$\_$6, PI Veilleux), where OH 119 $\mu$m is often seen
in emission but log $(L_{\rm AGN}$/L$_\odot)$ $<$ 12 (Mel\'endez et
al. 2013, in prep.), reinforces the idea that the AGN fraction is more
important in setting the character ({\em i.e.}  strength of emission
relative to absorption) of the OH feature than the AGN
luminosity. These results suggest that dominant AGN in ULIRGs and QSOs
provide a favorable environment to produce the OH molecule and excite
it (via radiative pumping or collisional excitation) to the
ground-state $^2\Pi_{3/2}$ $J$ = 5/2 level.  Interestingly, a similar
trend with AGN fraction was seen by Teng, Veilleux \& Baker (2013),
when examining the H~I 21-cm feature in a smaller set of {\em QUEST}
ULIRGs and QSOs with the Green Bank Telescope (GBT).

Figure 5 compares the strength of the OH 119 $\mu$m absorption feature
with that of the 9.7 $\mu$m silicate absorption feature, a measure of
obscuration in ULIRGs. The strengths of the 9.7 $\mu$m feature in our
sample galaxies are taken from two separate papers that use two
different methods to derive this quantity. In V09, $\tau_{\rm 9.7 um}$
is the effective 9.7 $\mu$m silicate optical depth measured relative
to the sum of the fitted blackbody components (a larger value of
$\tau_{\rm 9.7 um}$ implies a deeper silicate absorption feature).  In
Stierwalt et al. (2013), $S_{\rm 9.7 um}$ is the logarithm of the
ratio of the measured flux at the central wavelength of the silicate
absorption feature to the local continuum flux (a more negative value
of $S_{\rm 9.7 um}$ implies a deeper silicate absorption feature). In
both cases, we note that the more obscured systems ($\tau_{\rm 9.7
  um}$ $>$ 4 in V09 or $S_{\rm 9.7 um}$ $<$ --2 in Stierwalt et
al. 2013) more often show OH purely in absorption than the less
obscured systems. Conversely, systems with strong OH emission either
show very weak silicate absorption or silicate in emission. This last
category includes three of the four QSOs seen in emission in both OH
and silicate (F00509$+$1225 = I~Zw 1, PG~1440$+$356, and
PG~1613$+$658; PG~1351$+$640 is the only QSO that is not part of the
{\em Spitzer} sample of Schweitzer et al. 2008; these objects are not
plotted in Figure 5, but would reinforce the trend between FIR OH and
MIR silicate).

Figures 4 and 5 paint a consistent picture where both the OH 119
$\mu$m and 9.7 $\mu$m silicate features change character from
absorption to emission as the merger progresses and the AGN becomes
dominant.

\subsection{Distributions of Velocities and Wind Detection Rates}

Figure 6 shows the distributions of velocities derived from both the
OH absorption and emission line features [$v_{50}$(abs),
$v_{84}$(abs), $v_{50}$(emi), $v_{84}$(emi), and $v_{\rm max}$(abs),
as defined in \S 3.3 and listed in Table 2]. The velocity
distributions of the absorption features show a distinct excess at
negative values while the opposite is true of the emission
features. As mentioned in \S 3.3, all 15 objects with clear
composite absorption + emission features show a distinct P~Cygni
profile {\em i.e.} a blueshifted absorption feature accompanied with a
redshifted emission feature, an unambiguous signature of outflow.
As listed at the bottom of Table 2, the average (median)
  $v_{50}$(abs), $v_{84}$(abs), and $v_{\rm max}$(abs) are --194
  (--204), --444 (--492), and --927 (--925) km s$^{-1}$, respectively.
  Given possible projection effects, these measurements represent
  lower limits on the true velocities of the OH outflows in these
  objects.

Figures 6c and 6d show $v_{50}$(abs) and $v_{84}$(abs) in objects
  with pure absorption features and P Cygni profiles.  The average
  (median) $v_{50}$(abs) and $v_{84}$(abs) are $-$135 ($-$183) and
  $-$398 ($-$477) km~s$^{-1}$ for the objects with pure OH absorption
  profiles {\em versus} $-$262 ($-$243) and $-$495 ($-$495)
  km~s$^{-1}$ for the objects with P Cygni profiles. Given the
  measurement uncertainties ($\pm$50 km s$^{-1}$ or larger in some
  cases; Table 2), we do not consider these differences significant. A
  Kolmogorov-Smirnov (K-S) between these distributions supports this
  conclusion. As a case in point, we generally find excellent
  agreement between our velocities and those of S11 and
  Gonzalez-Alfonso et al. (2012; the largest discrepancy is seen for
  the P Cygni profile of F14387$-$3651, where $v_{84}$ in S11 is 240
  km s$^{-1}$ more blueshifted than the value reported
  here). Nevertheless, to remain aware of possible biases, the
  velocities derived in objects with OH P~Cygni profiles will be
  labeled differently from those derived in objects with OH purely in
  absorption or emission in all of the following figures.

We follow Rupke et al. (2005b) and conservatively define a wind as
having an OH absorption feature with a median velocity ($v_{50}$) more
negative than --50 km s$^{-1}$. Objects with OH purely in emission are
thus excluded by this definition, even though one of them
(PG~1613$+$658\footnote{F13451$+$1232 also shows broad OH profiles but
  this may partly be due to the binary nature of this object.}) shows
broad OH profiles extending in excess of $\sim$500 km s$^{-1}$,
suggestive of non-gravitational motion.\footnote{While winds produce
  blueshifted OH absorption profiles, they are not expected to produce
  shifts of the line profiles if OH is purely in emission, unless dust
  obscuration causes $\tau_{\rm FIR} >> 1$.} As described in Rupke et
al., the --50 km s$^{-1}$ cutoff is used to avoid contamination due to
systematic errors and measurement errors in wavelength calibration,
line fitting (see \S 3.3), and redshift determination. The few
redshifted absorption components detected in our sample have $v_{50}
\la$ 50 -- 100 km s$^{-1}$ (Figure 6), suggesting that this cutoff is
reasonable.

Using this conservative definition, we detect a molecular wind in 26
(70\%) of the 37 objects in our sample with measurable OH 119 $\mu$m
feature.  This fraction is $\sim$65\% and 62\% if the wind velocity
threshold is changed to $v_{50}$ = $-$75 and $-$100 km~s$^{-1}$,
respectively. The requirements to detect the wind in absorption
against the continuum source and for the absorption velocities to be
$\le$ --50 km s$^{-1}$ necessarily bias the sample against winds seen
edge-on. This wind detection rate is therefore a lower limit and our
results may be consistent with all ULIRGs having such molecular
winds. Rupke et al. (2005b) came to a similar conclusion regarding
neutral winds traced by Na I, although the sample of Rupke et
al. (2005b) is different from the present one in that it is made of
starburst dominated (U)LIRGs and therefore primarily traces
starburst-driven winds. The opening angle of the molecular
  outflows inferred for the wind detection rate of 70\% is
  $\sim$145$^\circ$, assuming all objects in our sample have an
  outflow.

Figure 7 shows no clear dependence of the wind detection rate on AGN
fraction, AGN luminosity, or star formation luminosity, to within the
uncertainties of the data  (recall that objects with OH feature in
  pure emission are excluded from this analysis; these objects are
  discussed in \S 4.7). In contrast, Rupke et al. (2005b)
detected Na~I winds slightly more frequently in ULIRGs ($\sim$75\%)
than in LIRGs ($\sim$45\%), and attributed this difference to higher
collimation of the outflows in LIRGs (see also Chen et al. 2010). The
results of Rupke et al. (2005b) refer to starburst dominated LIRGs,
which are not adequately sampled by our set of objects, so we cannot
verify if the dependence of the wind detection rate on star formation
rate seen in the neutral gas is also present in the molecular
gas. More relevant is the sample of Rupke et al. (2005c), which
contains several objects in common with the present sample. The lower
neutral wind detection rate found among Seyfert 2 ULIRGs in Rupke et
al. (2005c; 45\% $\pm$ 11\%) is not seen in OH, although the
comparison is necessarily limited to systems in the present sample
with $\alpha_{\rm AGN} \la$ 90\%.  We return to the multi-phase nature
of these winds in \S 5.

Only four objects in our sample (F12243$-$0036 = NGC~4418,
15250$+$3609, F17207$-$0014, and F22491$-$1808) show clear evidence
for inflow based on the detection of a redshifted OH absorption
feature with median velocity $v_{50}$(abs) $\ge$ 50 km s$^{-1}$. This
paucity suggests that the infalling gas, when present, generally
subtends a relatively small fraction of 4 $\pi$ steradians (e.g.,
planar or filamentary geometry). The fact that clear inverted P~Cygni
profiles are not seen in our sample of galaxies adds support to this
argument.  While Gonzalez-Alfonso et al. (2012) have shown that the
OH 119 $\mu$m feature in NGC~4418 is best fit as a weak inverted
P~Cygni profile, this conclusion is driven largely by the clear
detection of such a profile in [O~I] 63 $\mu$m. This result serves as
a cautionary tale that weak blueshifted OH emission may be missed in
our single-line fits.  Nevertheless, this does not change the fact
that inverted P~Cygni profiles with {\em strong} blueshifted emission
are absent from our sample of objects (although there are a few known
exceptions among other ULIRGs, e.g, Gracia Carpio et al. 2013, in prep).
This means that most of the infalling gas in ULIRGs lies in front of
the FIR continuum source and therefore does not subtend a large solid
angle as seen from the source.

\subsection{Kinematics {\em vs} Host Galaxy Masses and Merger Phases}

The velocities of local neutral gas outflows traced by Na I show a
tendency to become more negative with increasing host galaxy stellar
velocity dispersions and stellar masses (e.g., Rupke et al. 2002,
2005bc; Martin 2005; Chen et al. 2010). A similar trend is present
when using low-ionization tracers in $z \sim$ 1 star-forming galaxies
(Erb et al. 2012; c.f.\ Steidel et al. 2010; Law et al. 2012 at higher
redshifts). Figure 8 compares all of the velocities of the OH
absorption features derived from our data (regardless of whether they
indicate outflow or inflow) with these same host properties. We detect
no obvious trend with near-infrared derived stellar velocity
dispersions or stellar masses. Perhaps this is not too surprising
since our sample covers a considerably narrower range of host galaxy
properties than the Na I samples. Indeed, the low-$z$ trends between
Na I outflow velocities and host galaxy properties largely disappear
when only ULIRGs are considered. Moreover, Rothberg \& Fischer (2010)
and Rothberg et al. (2013) have recently argued that the near-infrared
derived velocity dispersions of luminous infrared merger remnants may
more closely reflect the properties of the young stellar disk than the
global properties of the host galaxy system as a whole. The noticeably
smaller stellar velocity dispersions among systems with OH P-Cygni or
pure-emission profiles is unexpected, and is not seen in the lower
panels of Figure 8 where the host stellar masses are considered. This
result may be related to the young stellar population bias in
near-infrared derived velocity dispersions noted by Rothberg \&
Fischer.

Since nearly all galaxies in our sample are undergoing a galaxy
interaction, it is natural to ask whether there is a dependence of the
molecular velocities on merger phase, or equivalently, ``interaction
class'' as defined in Veilleux, Kim, \& Sanders (2002). Given the
relatively small number of objects with detected OH absorption
features, we simplified the classification into ``binaries'', where
two distinct nuclei have been detected via arcsecond resolution
imaging, and ``singles'', where the two nuclei have coalesced. The
results are shown in Figure 9. The OH absorption velocities in
binaries are not statistically different from those in singles. This
conclusion is the same whether we use $v_{50}$(abs), $v_{84}$(abs),
and $v_{\rm max}$(abs). Recall, however, that IR-faint QSOs are
  not included in this analysis since their OH feature is in emission.
  One of the four IR-faint ULIRGs/QSOs with detected OH
  (PG~1613$+$658) shows broad OH emission profiles and may therefore
  qualify as having an outflow. We return to this point below (\S
  4.7).

\subsection{Kinematics {\em vs} Starburst Luminosities}

Plots of the velocities of local neutral gas outflows traced by Na I
(on the vertical axis) {\em versus} star formation rates (on the
horizontal axis) show an ``envelope'' with the most negative measured
outflow velocities increasing with star formation rates (e.g., Rupke
et al. 2005b; Martin 2005; Chen et al. 2010).  This trend seems driven
largely by the modest outflow velocities measured in dwarf galaxies
(Schwartz \& Martin 2004) and the large scatter in the outflow
velocities measured in the ULIRGs (Rupke et al. 2005b; Martin 2005). A
similar trend may be present at $z \ga$ 1 when considering wind
velocities derived from low-ionization species (e.g., Weiner et
al. 2009), although the situation likely is more complicated at these
higher redshifts (Steidel et al. 2010; Kornei et al. 2012; Martin et
al. 2012).  Figure 10 shows a similar plot substituting the OH
velocities for those based on Na I, covering $\sim$2 dex in
SFRs. Shown on the horizontal axis are the starburst luminosities
($\propto$ {\em SFR}) derived using $L_{\rm SB}$ $\equiv$ $(1 -
\alpha_{\rm AGN}) L_{\rm BOL}$ (see \S 2.2). No correlation, upper
envelope, or increase in scatter within the more luminous starbursts
is seen in this figure, regardless of which velocity [$v_{50}$(abs),
$v_{84}$(abs), or $v_{\rm max}$(abs)] is plotted.\footnote{Objects
  with significant inflows ($v_{50}$(abs) $\ge$ 50 km~s$^{-1}$) are
  excluded from this analysis, although this conclusion remains the
  same even when they are included.} If anything, there may even be a
tendency for the most luminous starbursts to have more positive
velocities (slower outflows), but this trend is not found to be
statistically significant: the probability, P[null], that the two
populations are drawn from the same parent population or are
uncorrelated is larger than $\sim$10\%, regardless of the statistical
test used for the analysis (see values listed in Table 3). Recall from
\S 2.2 that the ULIRGs with the most luminous starbursts in our sample
also tend to have smaller AGN fractions ($\la$ 40\%). As discussed in
the next section, these smaller AGN fractions might help explain the
slower outflows in these objects. Finally, we warn the readers that
this analysis is necessarily limited by the range in SFRs covered by
our sample ($\sim$2 dex), which is narrower than that in low-$z$ Na
I~D studies.

\subsection{Kinematics {\em vs} AGN Fractions}

Figure 11 shows the OH velocities plotted against the AGN fractions
$\alpha_{\rm AGN}$ derived from the $f_{15}/f_{30}$ ratios. A visual
inspection of this figure suggests that objects with dominant AGN
($\alpha_{\rm AGN} \ge$50\%) have more negative velocities (faster
outflows) than objects with dominant starbursts ($\alpha_{\rm AGN}
\le$50\%). However, a K-S test between the velocity distributions of
dominant AGN and dominant starburst systems indicates that this
difference is not statistically significant.\footnote{Once again,
  objects with significant inflows ($v_{50}$(abs) $\ge$ 50
  km~s$^{-1}$) are excluded from this analysis, although this
  conclusion remains the same even when they are included.} The
probability P[null] that the two populations are drawn from the same
parent population $\simeq$ 16\% for both $v_{50}$ and $v_{84}$
and $\sim$47\%  for $v_{\rm max}$ (Table 3). A similar negative
result is found when $\alpha_{\rm AGN}$ is varied from 40\% to 60\%
(beyond this range, fluctuations due to the small number of objects in
one of the two distributions dominate the statistics making the
results unreliable). However, a search for a correlation between the
OH kinematics and the AGN fraction produces more positive results with
P[null] $\simeq$ 1\% -- 5\% (not considering the more uncertain
$v_{\rm max}$; see Table 3). The fact that the Pearson's test for a
linear correlation produces small P[null] indicates that the data
favor a smooth transition with AGN fraction from low-velocity to
high-velocity outflows rather than a sudden threshold effect.  

The preceding analysis used an equal weighting scheme for all
  data points. The correlations with AGN fractions get slightly
  stronger if we remove from the analysis the velocities with the
  largest uncertainties (those indicated by small symbols in Figure 11
  and double colons in Table 2).

Note that a positive trend between wind velocities and AGN fractions
does not necessarily imply that the AGN is the driver of these
outflows. Once a significant fraction of the obscuring material has
been swept away, either by an AGN- or starburst-driven wind, we expect
to be able to see down the core of the ULIRG more easily. It therefore
becomes easier to observe the central high-velocity outflowing
material and detect the AGN via optical-MIR diagnostics. Orientation
effects may also be at play in some cases: both the AGN and central
high-velocity wind are more easily detectable in systems where the
dusty torus / disk are seen face-on. To test whether the AGN plays a
role in driving these outflows, we need to compare the outflow
kinematics with the AGN luminosities; this is done in the next
section.

\subsection{Kinematics {\em vs} AGN Luminosities}

Figure 12 displays the OH outflow velocities {\em versus} the AGN
luminosities, $L_{\rm AGN}$ $\equiv$ $\alpha_{\rm AGN} L_{\rm BOL}$
(\S 2.2). This figure shows distinctly more negative velocities in
systems with larger AGN luminosities. This trend is more evident with
$v_{50}$ and $v_{84}$ than with the more uncertain $v_{\rm max}$.
The null probability for a correlation between the OH velocities
  ($v_{50}$ or $v_{84}$) and log~$L_{\rm AGN}$ range from $\sim$0.4\%
  to $\sim$4\%, depending on the statistical test (Table 3). A visual
  inspection of panels $a$ and $b$ of Figure 12 also suggests a
  steepening of the relation between outflow velocities and
  log~$L_{\rm AGN}$ at high AGN luminosities: systems with $v_{50}
  \le$ --400 km~s$^{-1}$ all have AGN luminosities larger than ${\rm
    log}~(L_{\rm AGN}^{\rm break}/L_\odot) = 11.8 \pm 0.3.$ The
factor of 2 uncertainty on this luminosity break takes into account
the measurement uncertainty on this break visually estimated from the
panels of Figure 12 (estimated to be $\pm 0.2$) and the uncertainties
($\pm 0.3$) on the values of $\alpha_{\rm AGN}$ and $L_{AGN}$ for our
objects.  To verify the value of this break more quantitatively, K-S
tests were carried out between the velocity distributions of objects
with log $(L_{\rm AGN}/L_\odot) >$ $N$ and those with log $(L_{\rm
  AGN}/L_\odot) <$ $N$, where $N$ was varied from 11.7 to 12.0 (beyond
this range, fluctuations due to the small number of objects in one of
the two distributions dominate the statistics, making the results
unreliable). The results from these tests confirm that the most
significant difference between the two distributions occurs when $N$ =
11.8 (P[null] $\simeq$ 3\% for $v_{50}$ and 7\% for
$v_{84}$; Table 3). This break is not statistically significant when
using $v_{\rm max}$. {\em These results emphasize the need to use more
  robust values of the outflow velocities, such as $v_{50}$ and
  $v_{84}$, when carrying out these analyses. }

Once again, the preceding analysis used an equal weighting scheme
  for all data points. The conclusions remain the same if we remove
  from the analysis the velocities with the largest uncertainties
  (those indicated by small symbols in Figure 12 and double colons in
  Table 2), although in that case the significance of the correlations
  with AGN luminosities become virtually the same as with AGN
  fractions.

Recall that none of the four ULIRGs/QSOs with AGN fraction
$\alpha_{\rm AGN} \ga 90\%$ and detected OH 119 $\mu$m features,
including all of the OH-detected OT2 infrared-faint quasars, shows OH
absorption components. These objects thus do not appear in
  Figures 10 -- 12 and are excluded from our statistical analysis
  (Table 3).  The generally modest widths of the OH emission profiles
in these objects (PG~1613$+$658 is the only exception) do, however,
suggest that the winds, if present, have modest velocities. This
result seems inconsistent with the trend of increasing outflow
velocities with increasing AGN luminosities since all four objects
with AGN fraction $\alpha_{\rm AGN} \ga 90\%$ have AGN luminosities
near or above $10^{11.8}$ $L_\odot$. Perhaps we are seeing the end of
the fast wind phase predicted by some models (Narayanan et al. 2008;
Hopkins et al. 2009).

\section{Discussion}

Our kinetic analysis of the OH 119 $\mu$m velocities in nearby ULIRGs
and QSOs described in \S 4 shows trends of increasing molecular wind
velocities (becoming more negative) with increasing AGN fractions and
luminosities. While the former may be explained as an obscuration
effect where both the AGN and central high-velocity outflowing
material are more easily detectable when the dusty material has been
swept away or is seen more nearly face-on, the trend with AGN
luminosity suggests that the AGN is actually playing a role in driving
the fast winds.\footnote{We note, however, that for non-spherical
  geometries, even apparent AGN luminosities may be affected by
  optical depth and projection effects. For example, in the simple
  case of an optically thick ($\tau_{\rm FIR}$ $>>$ 1) torus
  surrounding a MIR-bright compact object, the luminosity can be
  significantly higher in the polar direction.} Our data favor the
existence of a non-linear (steepening) relation between outflow
velocities and the logarithms of the AGN luminosities. Above
\begin{eqnarray} {\rm log}~(L_{\rm AGN}^{\rm break}/L_\odot) = 11.8
  \pm 0.3,
\end{eqnarray}
AGN seems to play a dominant role in driving the outflowing molecular
gas in the objects of our sample. This luminosity break coincides
approximately with the minimum AGN luminosity traditionally used to
define a quasar ($M_B \le -23$). 

This conclusion only applies to objects with absorption signatures of
molecular outflows.  Fully coalesced ULIRG/QSO mergers with AGN
fractions $\alpha_{\rm AGN} \ga 90\%$, including all of the
OH-detected OT2 infrared-faint quasars where OH is seen purely in
emission, generally show modest line widths and thus modest outflows
despite their large AGN luminosities.

The results from S11 on 6 ULIRGs in common with our sample indicate
that the OH outflows are not only fast but also powerful, capable of
displacing a significant fraction of the entire molecular ISM of the
host galaxies. Given our methods of selection, there is no reason to
believe that the objects studied here are any different from those in
S11. Since the molecular ISM is the fuel for future star formation and
galaxy build-up, the quasar-driven outflows detected here may
therefore have a significant effect on the evolution of their hosts,
potentially quenching star formation in their centers. The modest OH
emission line widths in the infrared-faint quasars perhaps indicate
that AGN feedback subsides once the quasar has poked a substantial
hole through its natal ``cocoon'', although this is based on only four
objects.

There has been considerable theoretical support in recent years for
the idea of quenching star formation in galaxies via AGN-driven winds
(e.g., Narayanan et al. 2008; Hopkins et 2009 and references therein),
but relatively little direct observational evidence for it, until
recently (e.g., Feruglio et al. 2010; S11; RV11; RV13).  This last
paper also points out the possibility of increasing outflow velocities
at higher AGN luminosities. Using detailed IFU data on a sample of 6
ULIRGs, all from the {\em QUEST} sample, RV13 find that fast
AGN-driven neutral (Na I D) and/or ionized (H$\alpha$) winds are only
present in systems with log~$(L_{\rm AGN}^{\rm break}/L_\odot) \ge
11.7$.  This AGN luminosity, while only tentatively identified in that
paper since it is based on a very small sample, is consistent to
within the uncertainties to that inferred from the larger set of {\em
  Herschel} data. This suggests that the conditions to drive fast
neutral/ionized outflows are approximately the same as for the
molecular outflows.

In this context, it is instructive to compare the OH velocities
measured here with the Na I velocities measured by Rupke et
al. (2005a, 2005b, and 2005c) and Krug et al. (2013, in prep.) from
integrated spectra and RV13 from IFU data (Figure 13). Also shown in
this figure for comparison are the mid-infrared fine-structure line
outflow velocities deduced by Spoon \& Holt (2009) from {\em Spitzer}
integrated spectra, the H$\alpha$ velocities measured by RV13 from the
IFU data, and the H~I 21-cm velocities measured by Teng, Veilleux, \&
Baker (2013) from GBT integrated spectra. While the various velocities
are measured over different scale ($\la$200 pc for the OH velocities
according to S11 and $\la$1$-$2 kpc for the neutral and ionized gas
velocities), general trends are observed: The outflow velocities of
the molecular gas are often comparable to the velocities of the
neutral gas, but they are generally smaller (more positive) than the
velocities of the ionized gas (panels $b$ and $c$ in Figure 13). The
similarity between the OH and Na I velocities is particularly evident
in panel (b) of Figure 13 where the sometimes very large nuclear Na~I
and H$\alpha$ outflow velocities, produced on sub-pc scale (e.g.,
Mrk~231; Veilleux et al. 2013; Figure 13a), are excluded from the
means. This panel also reiterates the result found by RV13 that the
ionized gas is often, but not always, the fastest component of the
outflow. While dust obscuration may explain some of these variations
(especially for the highly ionized gas found near the center of
AGN-dominated ULIRGs; e.g., blue wings in [Ne~V] emission profiles of
Spoon \& Holt 2009), it likely does not explain all of the spatial
variations found in the IFU data.

These multi-phase comparisons may help constrain the processes
involved in accelerating the material entrained in the wind.  In most
cases, the acceleration $a = F/m$ is expected to depend strongly on
the ISM phase. For instance, in the case of ram pressure acceleration
of clouds in a fast wind or a jet (e.g., Strel'nitskii \& Sunyaev
1973; Chevalier \& Clegg 1985),
% (adding to the fact that
% known radio jets are rare in our targets; c.f. Teng, Veilleux, \&
% Baker 2013), 
the force on a spherical cloud of mass $M_{\rm cloud}$ and density
$\rho_{\rm cloud}$ is expected to scale as the area of the cloud or
$(M_{\rm cloud}/\rho_{\rm cloud})^{2/3}$ and thus we expect the
acceleration to scale as $(M_{\rm cloud} \rho_{\rm cloud}^2)^{-1/3}$
(we neglect gravity).  Naively, one would expect the molecular clouds
to be considerably denser and more massive, on average, than the
ionized gas clouds so one would expect the molecular material to
always move at lower velocities. This simple-minded argument may be
relaxed if each accelerated cloud has a multi-phase structure (e.g.,
ionized skin or halo surrounding a neutral/molecular core) and thus
contributes all at once to the molecular, neutral, and ionized phases
of the outflow.  But the significant difference in velocities often
seen between the ionized and molecular gas phases (Figure 13)
indicates that this situation cannot apply to the bulk of the
outflowing material since the difference in velocities between the
various phases would imply severe shearing/shredding and very short
cloud lifetimes [$\sim R_{\rm cloud} / \Delta v \sim 10^4~(R_{\rm
  cloud}/ {\rm pc}) (100~{\rm km~s}^{-1} / \Delta v)$ years, where
$R_{\rm cloud}$ is the cloud radius and $\Delta v$ is the difference
in velocities between the various gas phases].

Another possible source of acceleration for these winds is UV-IR
radiation pressure. It has the distinct characteristic to be less
efficient at accelerating ionized material than dusty
neutral/molecular material due to the different sources of opacity.
Within the dust sublimation radius, $R_{\rm subl} \sim L_{46}^{1/2}$
pc (where a dust sublication temperature $T_{\rm subl} \simeq 1200$ K
was assumed), the dominant source of opacity is electron scattering
and the corresponding luminosity condition to drive a (spherical) wind
is the well-known Eddington luminosity. Estimates of the black hole
masses $M_{\rm BH}$ in our sample galaxies range from $\sim$0.5 -- 5
$\times$ 10$^8$ $M_\odot$ (e.g, Vestergaard \& Peterson 2006; Veilleux
et al. 2009a), corresponding to Eddington luminosities $L_{\rm Edd}$ =
1.3 $\times$ 10$^{46}$ ($M_{BH} / 10^8$) = 0.65 -- 6.5 $\times$
10$^{46}$ ergs s$^{-1}$. Systems near the AGN luminosity break identified
in our data are therefore accreting at sub-Eddington values
$\Gamma \equiv L_{\rm AGN}^{\rm break} / L_{\rm Edd}$ $\sim$ 0.04 -- 0.4.

Outside the dust sublimation radius, the much larger dust opacity
prevails and UV-IR radiation pressure from the AGN may drive
significant winds.  In this situation where dust opacity dominates,
Murray, Quataert, \& Thompson (2005) have derived a handy expression
for the limiting, Eddington-like luminosity above which momentum
deposition from the quasar (and/or starburst) is presumed to be enough
to clear most of the gas out of galactic nuclei ({\em i.e.}  this is a
criterion to ``blow away'' the gas rather than just a ``blow out'' in
the nomenclature of De Young \& Heckman 1994):
\begin{eqnarray}
  L_M \simeq (4 f_g c)/G) \sigma^4 \simeq 3 \times 10^{46} (f_g / 0.1) (\sigma / 200~{\rm km~s}^{-1})^4~{\rm ergs~ s}^{-1}, 
\end{eqnarray}
where $f_g$ is the gas fraction of the host and $\sigma$ is the
stellar velocity dispersion, a measure of the depth of the
gravitational potential well. In our sample galaxies, $f_g \simeq 0.1$
on average, although there is considerable scatter among ULIRGs (e.g.,
Downes \& Solomon 1998; Combes et al. 2013 and references therein),
while $\sigma$ ranges from $\sim$120 to $\sim$280 km s$^{-1}$ with an
average value of $\sim$200 km s$^{-1}$ (see Figure 8; as mentioned in
\S 4.4, this value can be considered a lower limit, Rothberg \&
Fischer 2010; Rothberg et al. 2013).  The steep dependence of $L_M$ on
$\sigma$ implies that $L_{\rm AGN}^{\rm break}$ $\simeq$ 2 -- 100\% of
$L_M$, with an average value of $\sim$10\%, {\em i.e.}  the AGN
luminosity break above which the AGN plays a dominant role in driving
the outflowing molecular gas in local ULIRGs is generally lower than
the Eddington-like luminosity derived by Murray et al. (2005).

The fact that $L_{\rm AGN}^{\rm break}$ is lower than $L_M$ is not
surprising since the latter quantity is the {\em maximum} luminosity
an object may have before blowing away all of the gas from a galaxy
via radiation pressure on dust. In addition, the theoretically derived
$L_M$ is subject to a number of issues that may overestimate its value
in a real system: (1) First, it is important to note that the
expression for $L_M$ is an order-of-magnitude estimate of the actual
value. A number of simplifying assumptions were made to derive this
expression, e.g., the ISM is approximated as an isothermal sphere of
gas with constant gas fraction. The ISM in ULIRGs is undoubtedly much
more complex. Disky ISM distributions, such as those presumed to exist
in a number of ULIRGs (e.g., Downes \& Salomon 1998), offer less
resistance to winds along the disk rotation axis for a given $L_{\rm
  AGN}$. A good example is the wide-angle galactic wind in Mrk~231
(RV11, RV13), which is inferred to be oriented perpendicular to the
central molecular disk, yet it entrains a substantial fraction of the
gas in the nuclear region ({\em i.e.} it is closer to a ``blow away''
event than just a ``blow out'', so $L_M$ is relevant here). In-situ
formation of the accelerated clouds (Faucher-Gigu\`ere, Quataert, \&
Murray 2012) and/or energy-conserving acceleration due to inefficient
cooling in the shocked winds of AGN (Faucher-Gigu\`ere \& Quataert
2012) may also effectively reduce the value of $L_M$. (2) The
expression for the maximum luminosity only takes into account the
radiation pressure from the AGN. Momentum injection from loosely
collimated jets (e.g., Teng, Veilleux, \& Baker 2013) would reduce the
AGN luminosity needed for efficient molecular outflow. (3) While not
dominant in systems with AGN luminosities above $L_{\rm AGN}^{\rm
  break}$, the powerful starbursts known to exist in these systems
also contribute energy and momemtum to driving the winds (including
possibly significant contributions from cosmic rays). We have
neglected the starburst contributions in the present treatment. This
would relax the requirements on the AGN, effectively reducing $L_M$
from the AGN by a factor of up to $\sim$2.

Observationally, the gas mass fraction $f_g$ is difficult to determine
in ULIRGs and depends on the uncertain CO-to-H$_2$ conversion
factor. The value of $f_g$ likely varies considerably from one galaxy
to the other in our sample. This is particularly true of the
  IR-faint QSOs, although they are still undoubtedly gas-rich (e.g.,
%  Evans et al. 2001; 
  Scoville et al. 2003). Given the uncertainties on $f_g$ and
$\sigma$, the large galaxy-to-galaxy variations in these quantities,
and the steep dependence of $L_M$ on $\sigma$, it is in fact not
surprising to find that our data (Figure 12) favor a trend, albeit
non-linear, between OH velociites and AGN luminosities rather than a
clear-cut AGN luminosity threshold for powerful molecular outflows.
Moreover, the exact shape of this outflow velocity -- AGN luminosity
relation likely applies only to local gas-rich ULIRGs.  Comparisons
with other types of objects with different $f_g$, $\sigma$, and
redshifts have to be done with great care. The gas mass fraction of
local giant spirals is $\sim$7-10\% (Leroy et al. 2008; Saintonge et
al. 2011), similar to that of local ULIRGs, but there is mounting
evidence that it increases with redshifts, reaching $\sim$34 $\pm$ 5\%
and 44 $\pm$ 6\% among ``main-sequence'' galaxies at $z \sim1.2$ and
$\sim ~ 2.3$, respectively (Tacconi et al. 2010; Daddi et
al. 2010). New observations also indicate that the gas mass fraction
increases with redshift among ULIRGs, by a factor of 3 $\pm$ 1 from
$z$ = 0 to 1 (Combes et al. 2013). These results fit well with the
large gas mass fractions ($\sim$40\%) previously derived in $z \sim$ 2
-- 3.4 sub-mm galaxies (Tacconi et al. 2006). For objects with the
same stellar velocity dispersion, $L_M$ at high redshifts is thus
expected to be 3 -- 4 $\times$ larger than in local objects. Future
ALMA observations should be able to verify this assertion.

% SMGs Tacconi+06: f_g ~ 0.42, sigma_g = 100, v_c ~ 392 Table 2
% normal SF galaxies Tacconi+10 at z = 1.2 and 2.3 --> f_g = 34% and 44%
% Combes et al. 2013: f_g increase by factor of 3+/-1 from z=0 --> 1 in ULIRGs

\section{Conclusions}

{\em Herschel} PACS data from the SHINING guaranteed-time key project
were combined with data from cycles 1 and 2 open-time programs to
carry out a systematic search for molecular outflows in a sample of 43
nearby ULIRGs and QSOs using OH 119 $\mu$m. The main results from
this surveys are:

\begin{itemize}

\item The OH 119 $\mu$m feature was detected in 37 (86\%) of the 43
  sample galaxies.

\item The nature of the OH 119 $\mu$m feature (emission, absorption,
  or both) depends on the properties of the galaxies: OH emission is
  stronger relative to OH absorption in quasar-dominated systems,
  becoming completely dominant ({\em i.e.} pure OH emission) in
  objects where the quasar contributes more than $\sim$90\% to the
  bolometric luminosity (called the AGN fraction for short). This
  behavior is similar to that of the silicate 9.7 $\mu$m feature
  studied with {\em Spitzer} and the H~I 21-cm feature measured with
  GBT in a subset of ULIRGs and QSOs.

\item Unambiguous evidence for molecular outflows, based on the
  presence of absorption features with median velocities more
  blueshifted than --50 km s$^{-1}$, is seen in 26 (70\%) of the 37
  targets with detected OH 119 $\mu$m.  Given possible projection
    effects, this wind detection rate is a lower limit and is
  consistent with all ULIRGs having molecular winds with an average
  opening angle $\sim$145$^\circ$. In contrast, absorption features
  with median velocities above +50 km s$^{-1}$, indicative of inflow,
  are detected in only four objects, none of which show clear inverted
  P~Cygni profiles. This result suggests that molecular inflows
  subtend a relatively small fraction of 4 $\pi$ steradians (e.g.,
  planar or filamentary geometry).

\item Typical median outflow velocities are $\sim$ $-$200 km
  s$^{-1}$, but terminal outflow velocities of $\sim$ $-$1000 km
  s$^{-1}$ are detected in several objects. Deprojected outflow
    velocities will be even larger.  The kinematics of these
  molecular outflows do not show any obvious dependence on the
  properties of the host galaxies (e.g., near-infrared derived stellar
  velocity dispersions, stellar masses) and star formation rates,
  although our sample covers a range of properties that is narrower
  than that of low-$z$ non-molecular wind studies, where trends have
  been observed with these quantities.

\item There is a tendency for objects with dominant and luminous
    AGN to show more blueshifted OH velocities (faster outflows). Our
  data favor a steepening of the relation between outflow velocities
  and the logarithms of the AGN luminosities above log $(L_{\rm
    AGN}^{\rm break} / L_\odot) = 11.8 \pm 0.3$, although better
  statistics are needed, particularly at lower starburst and AGN
  luminosities, to confirm this non-linearity.  While the trend
  between outflow velocities and AGN fractions may be explained as an
  obscuration effect, where both the AGN and central high-velocity
  outflowing material are more easily detectable when the dusty
  material has been swept away or is seen more nearly face-on, the
  dependence of the velocities on the AGN luminosities suggests that
  the AGN is playing a role in driving the fast winds. Our
    results emphasize that the use of the terminal velocities as the
    sole measures of outflow velocities should be avoided due to the
    much larger uncertainties on these measurements.

\item The OH emission profiles in three of the four late-stage
  ULIRG/QSO mergers with AGN fractions above $\sim$90\% are narrow,
  despite their large AGN luminosities, suggesting that the winds in
  these objects, if present, often have modest velocities. Quasar
  feedback thus seems to subside among mergers that have cleared
    a path through the dusty circumnuclear material.
\end{itemize}

\acknowledgements We thank the referee for suggesting changes that
helped improve this paper. Support for this work was provided by NASA
through {\em Herschel} contracts 1427277 and 1454738 (S.V. and M.M.)
and contracts 1364043, 1435724, and 1456609 (J.F.).  S.V. also
acknowledges support from the Alexander von Humboldt Foundation for a
``renewed visit'' to Germany following up the original 2009 award, and
thanks the host institution, MPE Garching, where a portion of this
paper was written.  E.G-A is a Research Associate at the
  Harvard-Smithsonian Center for Astrophysics, and thanks the support
  by the Spanish Ministerio de Econom\'{\i}a y Competitividad under
  projects AYA2010-21697-C05-0 and FIS2012-39162-C06-01.  Basic
  research in IR astronomy at the Naval Research Laboratory is funded
  by the US Office of Naval Research. This research made use of {\em
  PySpecKit}, an open-source spectroscopic toolkit hosted at
http://pyspeckit.bitbucket.org.  This work has made use of NASA's
Astrophysics Data System Abstract Service and the NASA/IPAC
Extragalactic Database (NED), which is operated by the Jet Propulsion
Laboratory, California Institute of Technology, under contract with
the National Aeronautics and Space Administration.

\clearpage

\begin{deluxetable}{llllrllrlllll}
\tabletypesize{\scriptsize}
\rotate
\tablecolumns{13}
\tablewidth{0pc}
\tablecaption{Galaxy Properties}
\tablehead{
\colhead{Name}    & \colhead{Other} & \colhead{z}   & \colhead{$\frac{f_{15}}{f_{30}}$}    & \colhead{$\alpha_{AGN}$} &
\colhead{$\log {\rm L_{bol}}$}    & \colhead{$\log {\rm L_{SB}}$}   & \colhead{$\log {\rm L_{AGN}}$}    & \colhead{$\sigma_{*}$} & \colhead{M$_H$}    & \colhead{IC} & \colhead{Type}   & \colhead{Program ($t_{exp}$)}\\
\colhead{}    & \colhead{Name} & \colhead{}   & \colhead{}    & \colhead{(\%)} &
\colhead{($L_\odot$)}    & \colhead{($L_\odot$)}   & \colhead{($L_\odot$)}    & \colhead{(${\rm km~s^{-1}}$)}    & \colhead{(mag)} & \colhead{}   & \colhead{} &  \colhead{(hr)}\\
\cline{1-13}\\
\colhead{(1)}    & \colhead{(2)} & \colhead{(3)}   & \colhead{(4)}    & \colhead{(5)} &
\colhead{(6)}    & \colhead{(7)}   & \colhead{(8)}    & \colhead{(9)}    & \colhead{(10)} & \colhead{(11)}   & \colhead{(12)} &  \colhead{(13)}
}
\startdata
F00509+1225	&	I Zw 1	&	0.0610	&	0.434	&	90.1	&	12.01	&	11.00	&	11.96	&	188	&	\nodata	&	S	&	AGN 1	&	OT1 (1.3)	\\
F01572+0009	&	Mrk 1014	&	0.1631	&	0.161	&	64.6	&	12.68	&	12.23	&	12.49	&	200	&	-25.39	&	S	&	AGN 1	&	OT1 (1.0)	\\
F05024-1941	&		&	0.1920	&	0.050	&	7.3	&	12.43	&	12.40	&	11.30	&	\nodata	&	-24.49	&	B	&	AGN 2	&	OT1 (4.6)	\\
F05189-2524	&		&	0.0426	&	0.198	&	71.7	&	12.22	&	11.67	&	12.07	&	137	&	-23.96	&	S	&	AGN 2	&	GTO (0.5)	\\
07251-0248	&		&	0.0876	&	0.075	&	30.0	&	12.45	&	12.29	&	11.92	&	\nodata	&	\nodata	&	S	&	\nodata	&	GTO (1.1)	\\
F07599+6508	&		&	0.1483	&	0.377	&	87.6	&	12.59	&	11.68	&	12.53	&	\nodata	&	-24.55	&	S	&	AGN 1	&	OT1 (3.7)	\\
F08572+3915	&		&	0.0584	&	0.191	&	70.4	&	12.20	&	11.67	&	12.05	&	\nodata	&	-23.58	&	B	&	L	&	GTO (1.0)	\\
09022-3615	&		&	0.0596	&	0.125	&	54.9	&	12.35	&	12.00	&	12.09	&	\nodata	&	\nodata	&	S	&	\nodata	&	GTO (0.5)	\\
F09320+6134	&	UGC 5101	&	0.0394	&	0.130	&	56.4	&	12.05	&	11.69	&	11.80	&	\nodata	&	\nodata	&	S	&	L	&	GTO (0.3)	\\
F10565+2448	&		&	0.0431	&	0.105	&	47.1	&	12.10	&	11.82	&	11.77	&	\nodata	&	\nodata	&	B	&	H	&	GTO (0.5)	\\
F11119+3257	&		&	0.1890	&	0.266	&	79.9	&	12.71	&	12.02	&	12.61	&	\nodata	&	-23.44	&	S	&	AGN 1	&	OT1 (4.6)	\\
Z11598-0114	&		&	0.1507	&	0.210	&	73.5	&	12.60	&	12.03	&	12.47	&	\nodata	&	-25.06	&	S	&	AGN 1	&	OT1 (0.8)	\\
F12072-0444	&		&	0.1286\tablenotemark{a}	&	0.220	&	74.8	&	12.46	&	11.86	&	12.33	&	\nodata	&	-24.22	&	B	&	AGN 2	&	OT1 (0.8)	\\
F12112+0305 	&		&	0.0733	&	0.060	&	17.8	&	12.38	&	12.30	&	11.63	&	\nodata	&	-24.36	&	B	&	L	&	GTO (0.5)	\\
F12243-0036	&	NGC 4418	&	0.0071	&	0.128	&	55.7	&	11.06	&	10.71	&	10.81	&	\nodata	&	\nodata	&	S	&	AGN 2	&	GTO (0.2)	\\
F12265+0219	&	3C 273	&	0.1583	&	0.840	&	98.5	&	12.86	&	11.04	&	12.86	&	\nodata	&	-28.06	&	S	&	AGN 1	&	OT1 (1.0)	\\
F12540+5708	&	Mrk 231	&	0.0422	&	0.272	&	80.5	&	12.60	&	11.89	&	12.51	&	120	&	-24.52	&	S	&	AGN 1	&	GTO (0.3)	\\
13120-5453	&		&	0.0308	&	0.080	&	33.4	&	12.30	&	12.12	&	11.83	&	\nodata	&	\nodata	&	S	&	AGN 2	&	GTO (0.3)	\\
F13305-1739	&		&	0.1484	&	0.377	&	87.5	&	12.32	&	11.42	&	12.26	&	\nodata	&	-25.47	&	S	&	AGN 2	&	OT1 (4.9)	\\
F13428+5608	&	Mrk 273	&	0.0378	&	0.081	&	34.2	&	12.21	&	12.03	&	11.74	&	285	&	-24.32	&	S	&	AGN 2	&	GTO (0.3)	\\
F13451+1232	&		&	0.1217	&	0.273	&	80.6	&	12.38	&	11.67	&	12.29	&	146,167\tablenotemark{c}	&	-25.52	&	B	&	AGN 2	&	OT1 (1.0)	\\
F14348-1447	&		&	0.0830	&	0.060	&	17.4	&	12.40	&	12.31	&	11.64	&	160	&	-24.99	&	B	&	L	&	GTO (1.1)	\\
F14378-3651	&		&	0.0676	&	0.064	&	21.1	&	12.17	&	12.07	&	11.50	&	153	&	\nodata	&	S	&	AGN 2	&	GTO (1.1)	\\
F14394+5332	&		&	0.1045	&	0.152	&	62.5	&	12.17	&	11.74	&	11.96	&	\nodata	&	-24.88, -23.56\tablenotemark{d}	&	B\tablenotemark{c}	&	AGN 2
&	OT1 (1.0)	\\
F15206+3342	&		&	0.1244	&	0.158	&	63.9	&	12.30	&	11.86	&	12.10	&	\nodata	&	-24.47	&	S	&	H	&	OT1 (1.5)	\\
F15250+3608	&		&	0.0552	&	0.095	&	42.3	&	12.10	&	11.86	&	11.73	&	150	&	\nodata	&	S	&	L	&	GTO (0.5)	\\
F15327+2340	&	Arp 220	&	0.0181	&	0.049	&	5.8	&	12.22	&	12.19	&	10.98	&	164	&	\nodata	&	B	&	L	&	GTO (0.9)	\\
F15462-0450	&		&	0.0998	&	0.145	&	60.6	&	12.27	&	11.86	&	12.05	&	169	&	-23.4	&	S	&	AGN 1	&	OT1 (1.0)	\\
F16504+0228	&	NGC 6240	&	0.0245	&	0.130	&	56.2	&	11.90	&	11.54	&	11.65	&	229	&	\nodata	&	B	&	L	&	GTO (0.3)	\\
F17207-0014	&		&	0.0428	&	0.038	&	$\le$ 5.0	&	12.45	&	12.43	&	 $\le$ 11.15	&	229	&	\nodata	&	S	&	L	&	GTO (0.3)	\\
F19297-0406	&		&	0.0857	&	0.066	&	23.4	&	12.44	&	12.33	&	11.81	&	\nodata	&	\nodata	&	S	&	H	&	GTO (0.3)	\\
19542+1110	&		&	0.0624\tablenotemark{b}	&	0.069	&	25.5	&	12.12	&	11.99	&	11.52	&	\nodata	&	\nodata	&	S	&	L	&	GTO (1.1)	\\
F20551-4250	&		&	0.0430	&	0.132	&	56.9	&	12.11	&	11.74	&	11.87	&	140	&	\nodata	&	S	&	L	&	GTO (0.5)	\\
F22491-1808  	&		&	0.0778	&	0.057	&	14.4	&	12.90	&	12.83	&	12.05	&	\nodata	&	-24.21	&	B	&	H	&	GTO (0.8)	\\
F23128-5919 	&	ESO 148-IG 002	&	0.0446	&	0.154	&	63.0	&	12.09	&	11.66	&	11.89	&	\nodata	&	\nodata	&	B	&	H	&	GTO (0.3)	\\
F23233+2817	&		&	0.1140	&	0.204	&	72.5	&	12.11	&	11.55	&	11.97	&	\nodata	&	-24.54	&	S	&	AGN 2	&	OT1 (1.0)	\\
F23365+3604	&		&	0.0645	&	0.100	&	44.6	&	12.22	&	11.97	&	11.87	&	145	&	\nodata	&	S	&	L	&	GTO (0.5)	\\
F23389+0300	&		&	0.1450	&	0.066	&	22.7	&	12.19	&	12.07	&	11.54	&	\nodata	&	-24.15	&	B	&	AGN 2	&	OT1 (4.4)	\\
PG 1126-041 	&	Mrk 1298	&	0.0620	&	0.329	&	88.7	&	11.52	&	10.57	&	11.47	&	194	&	-24.15	&	S	&	AGN 1	&	OT2 (3.1)	\\
PG 1351+640                   	&		&	0.0882	&	0.389	&	92.2	&	12.04	&	10.93	&	12.00	&	\nodata	&	\nodata	&	S	&	AGN 1	&	OT2 (10.6)	\\
PG 1440+356	&	Mrk 0478        	&	0.0780	&	0.361	&	90.7	&	11.80	&	10.77	&	11.76	&	\nodata	&	-24.65	&	S	&	AGN 1	&	OT2 (5.4)	\\
PG 1613+658	&	Mrk 0876            	&	0.1290	&	0.398	&	92.6	&	12.29	&	11.16	&	12.26	&	\nodata	&	-25.68	&	S	&	AGN 1	&	OT2 (4.1)	\\
PG 2130+099 	&	UGC 11763         	&	0.0630	&	0.563	&	98.5	&	11.77	&	9.95	&	11.76	&	156	&	-24.14	&	S	&	AGN 1	&	OT2 (14.0)	\\
\enddata
%% Text for table notes should follow after the \enddata but before
%% the \end{deluxetable}. Make sure there is at least one \tablenotemark
%% in the table for each \tablenotetext.

\tablecomments{Column 1: galaxy name. Coordinate-based names beginning
  with “F” are sources in the IRAS Faint Source Catalog. Column 2:
  another name. Column 3: redshift. Column 4: $\frac{f_{15}}{f_{30}}$
  values from V09 and Stierwalt et al. (2013). When no 15-to-30
  $\micron$ data is available the value is estimated by using the
  $\frac{f_{25}}{f_{60}}$ ratio and the good correlation between the
  25-to-60 vs. 30-to-15 \micron flux ratios (see V09). Column 5:
  $\alpha_{AGN}$, fractional contribution of the AGN to the bolometric
  luminosity based on the $\frac{f_{15}}{f_{30}}$. See Section
  2.2. Column 6: bolometric luminosity. See Section 2.2 for more
  detail. Column 7: starburst bolometric luminosity. See Section 2.2
  for more detail. Column 7: AGN luminosity. See Section 2.2 for more
  detail. Column 8: $\sigma_*$, stellar velocity dispersion from
  Dasyra et al. (2006a, 2006b, 2007). Column 10: M$_H$, absolute
  H-band magnitudes from Veilleux et al. (2006, 2009b). For the sake of
  completeness we also included M$_H$ absolute magnitudes derived from
  M$_{k'}$ values assuming H-K' = 0.35 mag (Veilleux et
  al. 2002). Column 11: Interaction Class, "S" for singles and "B" for
  binaries, based on classifications of Veilleux et
  al. (2002, 2006, 2009b) and, when not available, our own visual
  classification from HST-ACS F814W images (Iwasawa et
  al. 2011). Column 12: optical spectral types of our sample, HII (for
  HII galaxies), L (for LINER-like), AGN 2 and AGN 1 (for type 2 AGN
  and type 1 AGN, respectively). For the spectral type we adopted the
  classification from Veilleux et al. (1999) and V09, when not available,
  we used the values from NED/SIMBAD. Column 13: program, GTO SHINING
  program (PI: Sturm, E), QUEST OT1 and OT2 programs (PI: Veilleux,
  S). In parenthesis is the total PACS exposure time. }

\tablenotetext{a}{Redshift from Spoon et al. (2009). This value is
  based on [Ne~II] 12.8 $\mu$m and is slightly larger than that derived
  from optical data (e.g. Strauss et al. 1992).
% systemic velocitiy used in this
%  figure (38,550 km~s$^{-1}$; Spoon \& Holt 2009) is derived from
%  [Ne~II] 12 $\mu$m and is slightly larger than that derived from
%  optical data (38,480 km~s$^{-1}$; Strauss et al. 1992). }
}

\tablenotetext{b}{Average redshift derived from our fit to the
  [Ne~II], [Ne~III] and [O~IV] emission lines in the {\it Spitzer} IRS
  spectra.}
\tablenotetext{c}{The values presented correspond to the eastern and western nucleus of F13451+1232, respectively.}
\tablenotetext{d}{F14394$+$5332 is a triple system with an eastern pair and a
  western component.}

\end{deluxetable}

\begin{deluxetable}{lrrrrrrrrrr}
\tabletypesize{\scriptsize}
\rotate
\tablecolumns{11}
\tablewidth{0pc}
\tablecaption{Properties of the OH 119 $\mu$m Profiles}
\tablehead{
\colhead{(Name)}    & \colhead{v$_{50}$ (abs)} & \colhead{v$_{84}$ (abs)}   & \colhead{v$_{\rm max}$ (abs)} & \colhead{Flux$_{\rm abs}$} & \colhead{EQW$_{\rm abs}$}    & \colhead{v$_{50}$ (emi)} &
\colhead{v$_{84}$ (emi)}    & \colhead{Flux$_{\rm emi}$}    & \colhead{EQW$_{\rm emi}$}   &  \colhead{EQW$_{\rm Total}$}\\
\colhead{}    & \colhead{km s$^{-1}$} & \colhead{km s$^{-1}$} & \colhead{km s$^{-1}$}   & \colhead{Jy km s$^{-1}$}    & \colhead{km s$^{-1}$} &
\colhead{km s$^{-1}$}    & \colhead{km s$^{-1}$}   & \colhead{Jy km s$^{-1}$}    & \colhead{km s$^{-1}$}    & \colhead{km s$^{-1}$}\\
\cline{1-11}\\
\colhead{(1)}    & \colhead{(2)} & \colhead{(3)}   & \colhead{(4)}    & \colhead{(5)} &
\colhead{(6)}    & \colhead{(7)}   & \colhead{(8)}    & \colhead{(9)}    & \colhead{(10)} & \colhead{(11)}
}
\startdata
F00509+1225	&	\nodata	&	\nodata	&	\nodata	&	\nodata	&	\nodata	&	171	&	363	&	150.0	&	-65	&	-65	\\
F01572+0009	&	-724::	&	-892::	&	-1100::	&	-32.8::	&	23::	&	195::	&	345::	&	32.7::	&	-24::	&	0::	\\
F05024-1941	&	-183	&	-508::	&	-850::	&	-83.6	&	83	&	\nodata	&	\nodata	&	\nodata	&	\nodata			&	83	\\
F05189-2524	&	-327	&	-574	&	-850	&	-510.0	&	59	&	189	&	351	&	424.1	&	-51	&	10	\\
07251-0248	&	-63	&	-255	&	-550	&	-453.8	&	56	&	\nodata	&	\nodata	&	\nodata	&	\nodata			&	56	\\
F07599+6508	&	-459	&	-652	&	-1000	&	-47.7	&	35	&	315	&	808	&	74.1	&	-53	&	-19	\\
F08572+3915	&	-489	&	-832	&	-1100	&	-347.5	&	124	&	\nodata	&	\nodata	&	\nodata	&	\nodata			&	124	\\
09022-3615	&	-153	&	-297	&	-650	&	-265.5	&	33	&	195	&	339	&	405.7	&	-51	&	-17	\\
F09320+6134	&	-9	&	-225	&	-1200	&	-5.0	&	83	&	\nodata	&	\nodata	&	\nodata	&	\nodata			&	83	\\
F10565+2448	&	-267	&	-489	&	-950	&	-2140.7	&	156	&	99	&	273	&	1099.2	&	-82	&	75	\\
F11119+3257	&	-423	&	-814:	&	-1200:	&	-141.6	&	124	&	\nodata	&	\nodata	&	\nodata	&	\nodata			&	124	\\
Z11598-0114	&	-153	&	-477	&	-800	&	-295.2	&	121	&	\nodata	&	\nodata	&	\nodata	&	\nodata			&	121	\\
F12072-0444	&	-69	&	-321	&	-1200	&	-221.9	&	135	&	243	&	393	&	137.6	&	-87	&	51	\\
F12112+0305 	&	-117	&	-237	&	-400	&	-0.8	&	52	&	243	&	447	&	0.8	&	-48	&	2	\\
F12243-0036	&	111	&	-21	&	\nodata	&	-1406.6	&	63	&	\nodata	&	\nodata	&	\nodata	&	\nodata			&	63	\\
F12265+0219	&	\nodata	&	\nodata	&	\nodata	&	\nodata	&	\nodata			&	\nodata	&	\nodata	&	\nodata	&	\nodata			&	\nodata			\\
F12540+5708	&	-237	&	-610	&	-1500	&	-5510.7	&	243	&	147	&	357	&	2956.9	&	-134	&	113	\\
13120-5453	&	-195	&	-520	&	-1200	&	-5037.9	&	113	&	\nodata	&	\nodata	&	\nodata	&	\nodata			&	113	\\
F13305-1739	&	\nodata	&	\nodata	&	\nodata	&	\nodata	&	\nodata			&	\nodata	&	\nodata	&	\nodata	&	\nodata			&	\nodata			\\
F13428+5608	&	-201	&	-495	&	-750	&	-679.2	&	61	&	\nodata	&	\nodata	&	\nodata	&	\nodata			&	61	\\
F13451+1232	&	\nodata	&	\nodata	&	\nodata	&	\nodata	&	\nodata			&	-51	&	190	&	203.2	&	-136	&	-136	\\
F14348-1447	&	-291:	&	-508	&	-900	&	-696.1	&	112	&	93	&	273	&	542.1	&	-88	&	25:	\\
F14378-3651	&	-219	&	-556	&	-1200	&	-1009.4	&	206	&	261	&	423	&	423.6	&	-91	&	119	\\
F14394+5332	&	-291	&	-495	&	-750    &	-106.9	&	62	&	387	&	520	&	43.0	&	-26	&	37	\\
F15206+3342	&	\nodata	&	\nodata	&	\nodata	&	\nodata	&	\nodata			&	\nodata	&	\nodata	&	\nodata	&	\nodata			&	\nodata			\\
F15250+3608	&	189	&	-21	&	\nodata		&	-491.6	&	125	&	\nodata	&	\nodata	&	\nodata	&	\nodata			&	125	\\
F15327+2340	&	21	&	-153	&	-700	&	-22450.3	&	211	&	\nodata	&	\nodata	&	\nodata	&	\nodata			&	-211	\\
F15462-0450	&	-225:	&	-459:	&	-600:	&	-161.9	&	80:	&	\nodata	&	\nodata	&	\nodata	&	\nodata			&	80:	\\
F16504+0228	&	-207	&	-544	&	-1200	&	-2126.8	&	84	&	\nodata	&	\nodata	&	\nodata	&	\nodata			&	84 \\
F17207-0014	&	51	&	-165	&	\nodata		&	-4619.4	&	148	&	\nodata	&	\nodata	&	\nodata	&	\nodata			&	148	\\
F19297-0406	&	-231	&	-532	&	-1000	&	-750.5	&	119	&	\nodata	&	\nodata	&	\nodata	&	\nodata			&	119	\\
19542+1110	&	-93:	&	-489:	&	-700	&	-331.9	&	69	&	243	&	441	&	191.2	&	-41 &	29:	\\
F20551-4250	&	-381	&	-748	&	-1200	&	-2.1	&	70	&	\nodata	&	\nodata	&	\nodata	&	\nodata			&	70	\\
F22491-1808  	&	99	&	3::	&	\nodata	&	-0.3	&	25:	&	\nodata	&	\nodata	&	\nodata	&	\nodata			&	25:	\\
F23128-5919 	&	\nodata	&	\nodata	&	\nodata	&	\nodata	&	\nodata			&	\nodata	&	\nodata	&	\nodata	&	\nodata			&	\nodata			\\
F23233+2817	&	-267	&	-423	&	-500	&	-41.3	&	30	&	201	&	339	&	79.2	&	-58	&	-27	\\
F23365+3604	&	-243	&	-604	&	-1300	&	-1348.8	&	197	&	153	&	321	&	619.1	&	-94	&	107	\\
F23389+0300	&	-171	&	-285	&	-600	&	-43.4	&	48	&	351	&	634	&	86.5	&	-91	&	-46	\\
PG 1126-041 	&	\nodata	&	\nodata	&	\nodata	&	\nodata	&	\nodata			&	\nodata	&	\nodata	&	\nodata	&	\nodata			&	\nodata			\\
PG 1351+640     &	\nodata	&	\nodata	&	\nodata	&	\nodata	&	\nodata			&	 33	&	171	&	12.8	&	-40	&	-40		\\
PG 1440+356	&	\nodata	&	\nodata	&	\nodata	&	\nodata	&	\nodata			&	-105	&	69	&	31.1	&	-43	&	-43	\\
PG 1613+658	&	\nodata	&	\nodata	&	\nodata	&	\nodata	&	\nodata			&	135	&	333	&	55.8	&	-80	&	-80	\\
PG 2130+099 	&	\nodata	&	\nodata	&	\nodata	&	\nodata	&	\nodata			&	\nodata	&	\nodata	&	\nodata	&	\nodata			&	\nodata			\\
\hline
\hline
{\bf Average} & -194 &  -444 &  -927 & -1605 &  103 &  202 & 401  &  439 &  -72 &67 \\

{\bf Median}  & -204 &  -492 &  -925 & -340 &  84 &   198 & 354  & 197  & -70&73\\

\enddata
%% Text for table notes should follow after the \enddata but before
%% the \end{deluxetable}. Make sure there is at least one \tablenotemark
%% in the table for each \tablenotetext.

\tablecomments{The uncertainties on the equivalent widths and fluxes
  (velocities, except $v_{\rm max}$) are typically 20\% (50 km
  s$^{-1}$), unless the value is followed by a colon (when the
  uncertainties are 20-50$\%$ and 50-150 km s$^{-1}$, respectively) or
  a double colon (when the uncertainties are larger than 50$\%$ and
  150 km s$^{-1}$, respectively). For $v_{\rm max}$, the typical
  uncertainty is $\pm$200 km s$^{-1}$. Column 1: galaxy
  name. Coordinate-based names beginning with “F” are sources in the
  IRAS Faint Source Catalog. Column 2: $v_{50}$(abs) is the median
  velocity of the fitted absorption profile {\em i.e.} 50\% of the
  absorption takes place at velocities above - more positive than -
  $v_{50}$. Column 3: $v_{84}$(abs) is the velocity above which 84\%
  of the absorption takes place. Column 4: $v_{\rm max}$(abs) is the
  maximum extent of the blueshifted wing of the OH 119.233 $\mu$m
  absorption profile. Column 5: the total integrated flux for the
  absorption component(s). Column 6: the total equivalent width for
  the absoprtion component(s). Column 7: $v_{50}$ (emi) is the median
  velocity of the fitted emission profile. Column 8: $v_{84}$ (emi) is
  the velocity below which 84\% of the emission takes place. Column 9:
  the total integrated flux for the emission component(s). Column 10:
  the total equivalent width for the emission component(s). Column 11:
  The total equivalent width for the sum of the two components for one
  line of the OH doublet.}

\end{deluxetable}

\begin{deluxetable}{lcrrrlrlll}
\tabletypesize{\scriptsize}
\rotate
\tablecolumns{10}
\tablewidth{0pc}
\tablecaption{Results from the Statistical Analysis for Sources with v$_{50}$(abs) $<$ 50 km s$^{-1}$ }
\tablehead{
\colhead{Parameter}&  \colhead{\# of Objects}& \colhead{Break}&  \colhead{P$_{K-S}$}    &   \colhead{$\rho_s$}    &  \colhead{P$_\rho$}    &  \colhead{$\tau$}    &  \colhead{P$_\tau$}    &  \colhead{r}    &  \colhead{P$_r$}    \\
\cline{1-10}\\
\colhead{(1)}    & \colhead{(2)} & \colhead{(3)}   & \colhead{(4)}    & \colhead{(5)} & \colhead{(6)}    & \colhead{(7)} & \colhead{(8)}   & \colhead{(9)} & \colhead{(10)}
}
\startdata

$\log {\rm L_{SB}}$--v$_{50}$	&	28& 12	&	0.11	&	0.33	&	0.087	&	0.23	&	0.086	&	0.18	&	0.37	\\
$\log {\rm L_{SB}}$--v$_{84}$ 	&	28&	12	&	0.54	&	0.19	&	0.33	&	0.14	&	0.30	&	0.19	&	0.32	\\
$\log {\rm L_{SB}}$--v$_{\rm max}$	&	28&	12	&	0.54	&	0.21	&	0.27	&	0.15	&	0.25	&	0.23	&	0.25	\\
&       & \\
 $\alpha_{AGN}$--v$_{50}$	&	28&	50	&	0.16	&	-0.47	&	{\bf 0.012}	&	-0.34	&	{\bf 0.012}	&	-0.47	&	{\bf 0.011}	\\
 $\alpha_{AGN}$--v$_{84}$	&	28&	50	&	0.16	&	-0.39	&	{\bf 0.038}	&	-0.26	&	{\bf 0.049}	&	-0.45	&	{\bf 0.015}	\\
$\alpha_{AGN}$--v$_{\rm max}$	&	28&	50	&	0.47	&	-0.32	&	0.10	&	-0.22	&	0.097	&	-0.35	&	0.069	\\
&       & \\
$\log {\rm L_{AGN}}$--v$_{50}$	&	28&	11.8	&	{\bf 0.032}	&	-0.46	&	{\bf 0.013}	&	-0.34	&	{\bf 0.012}	&	-0.53	&	{\bf 0.0036}	\\
$\log {\rm L_{AGN}}$--v$_{84}$	&	28&	11.8	&	0.071	& -0.39	&	{\bf 0.038}	&	-0.27	&	{\bf 0.041}	&	-0.50	&	{\bf 0.0070}	\\
$\log {\rm L_{AGN}}$--v$_{\rm max}$	&	28&	11.8	&	0.44	&	-0.24	&	0.23	&	-0.19	&	0.16	&	-0.33	&	0.087	\\

\enddata
 
 \tablecomments{Column 1. Quantities considered for the
  statistical
 test. Column 2. Number of objects with no significant
  inflows [$v_{50}$(abs) $<$ 50 km s$^{-1}$]. Column 3. Value of the
  break used to define the two
 distributions for the
  Kolmogorov-Smirnov (K-S) test. Column 4. Null
 probability of the
  K-S test. Column 5. Spearman rank order
 correlation coefficient.
  Column 6. Null probability of the Spearman
 rank order
  correlation. Column 7.  Kendall's correlation
 coefficient. Column
  8. Null probability of Kendall's
 correlation. Column 9. Pearson's
  linear correlation
 coefficient. Column 10. Two-tail area
  probability of Pearson's linear
 correlation. Null probabilities
  less than $\sim$0.05 (shown in
 bold-faced characters) indicate
  statistically significant trends; those with null probabilities
  between 0.05 and 0.10 are tentative.}
 
\end{deluxetable}

\begin{deluxetable}{lcc}
\tabletypesize{\scriptsize}
\rotate
\tablecolumns{9}
\tablewidth{0pc}
\tablecaption{Parameters from Linear Regression Analysis for Sources with v$_{50}$(abs) $<$ 50 km s$^{-1}$ }
\tablehead{
\colhead{X-Y}&    \multicolumn{2}{c}{$Y = aX + b$}   \\
\cline{1-3}\\
   & \colhead{$a$} & \colhead{$b$}  
}
\startdata

 $\alpha_{AGN}$--v$_{50}$	&	-3.0	$\pm$	1.0	&	-92.0	$\pm$	40.3	\\
 $\alpha_{AGN}$--v$_{84}$	&	-3.5	$\pm$	1.2	&	-332.7	$\pm$	61.5	\\
$\alpha_{AGN}$--v$_{\rm max}$\tablenotemark{a}	&	-4.0	$\pm$	1.9	&	-732.0	$\pm$	98.4	\\
&       & \\
$\log {\rm L_{AGN}}$--v$_{50}$	&	-212	$\pm$	72	&	2286	$\pm$	838	\\
$\log {\rm L_{AGN}}$--v$_{84}$	&	-236	$\pm$	73	&	2310	$\pm$	867	\\
$\log {\rm L_{AGN}}$--v$_{\rm max}$\tablenotemark{a}	&	-237	$\pm$	100	&	1891	$\pm$	1185	\\

\enddata
 
 \tablecomments{$X$ and $Y$ represent the independent and
  dependent variables, respectively. $a$ and $b$ represent the regression
  coefficient (slope) and regression constant (intercept),
  respectively. For the relationship presented in this table we used
  the ordinary least-square regression of the dependent variable, $Y$,
  against the independent variable $X$, OLS($Y$|$X$)
 }

\tablenotetext{a}{Note from Table 3 that these correlations are
  tentative so the parameters listed here should be treated with
  caution.}

\end{deluxetable}

\clearpage

\begin{figure*}
\epsscale{1.1}
\plotone{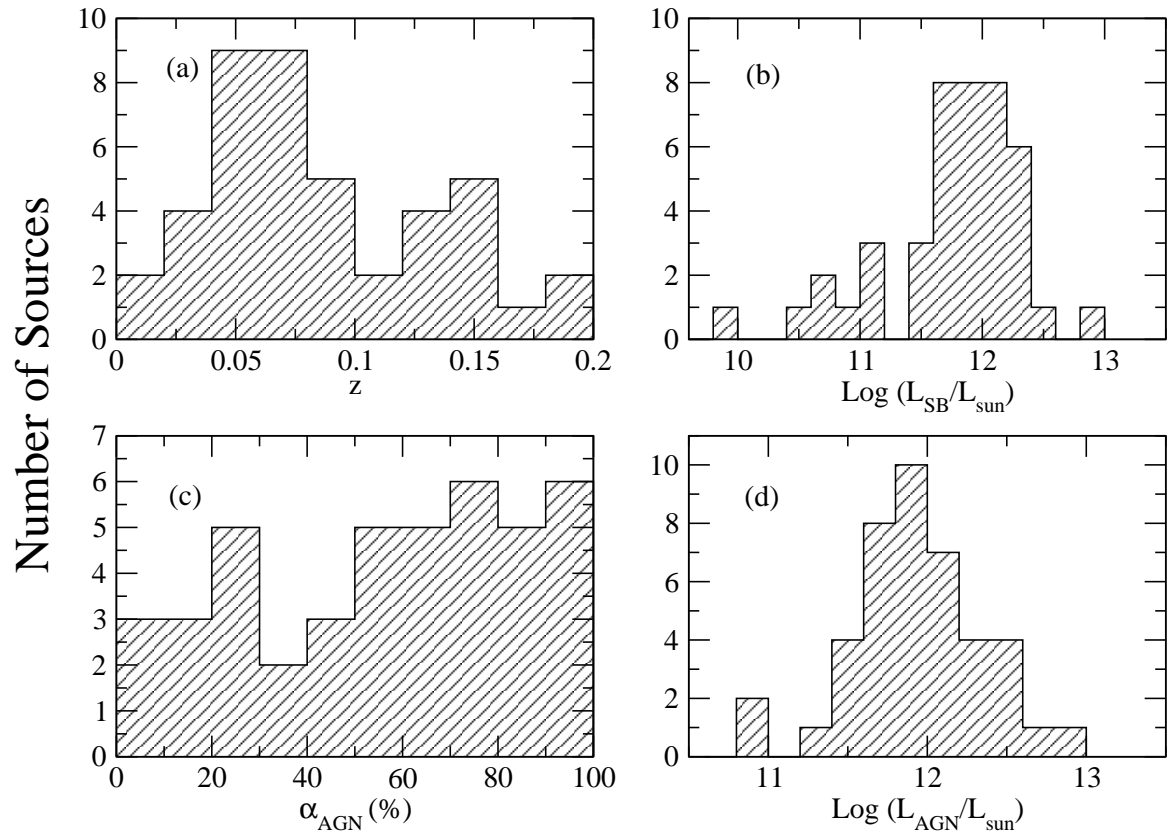}
% \centering
%\includegraphics[width=1.1\textwidth,angle=180]{fig1.eps}
\caption{ Histograms showing the distributions of the sample
  properties: ($a$) redshifts, ($b$) starburst luminosities, ($c$) AGN
  fractions, and ($d$) AGN luminosities.  }
%\label{fig:tmp}
\end{figure*}

\begin{figure*}
\epsscale{1.0}
\plotone{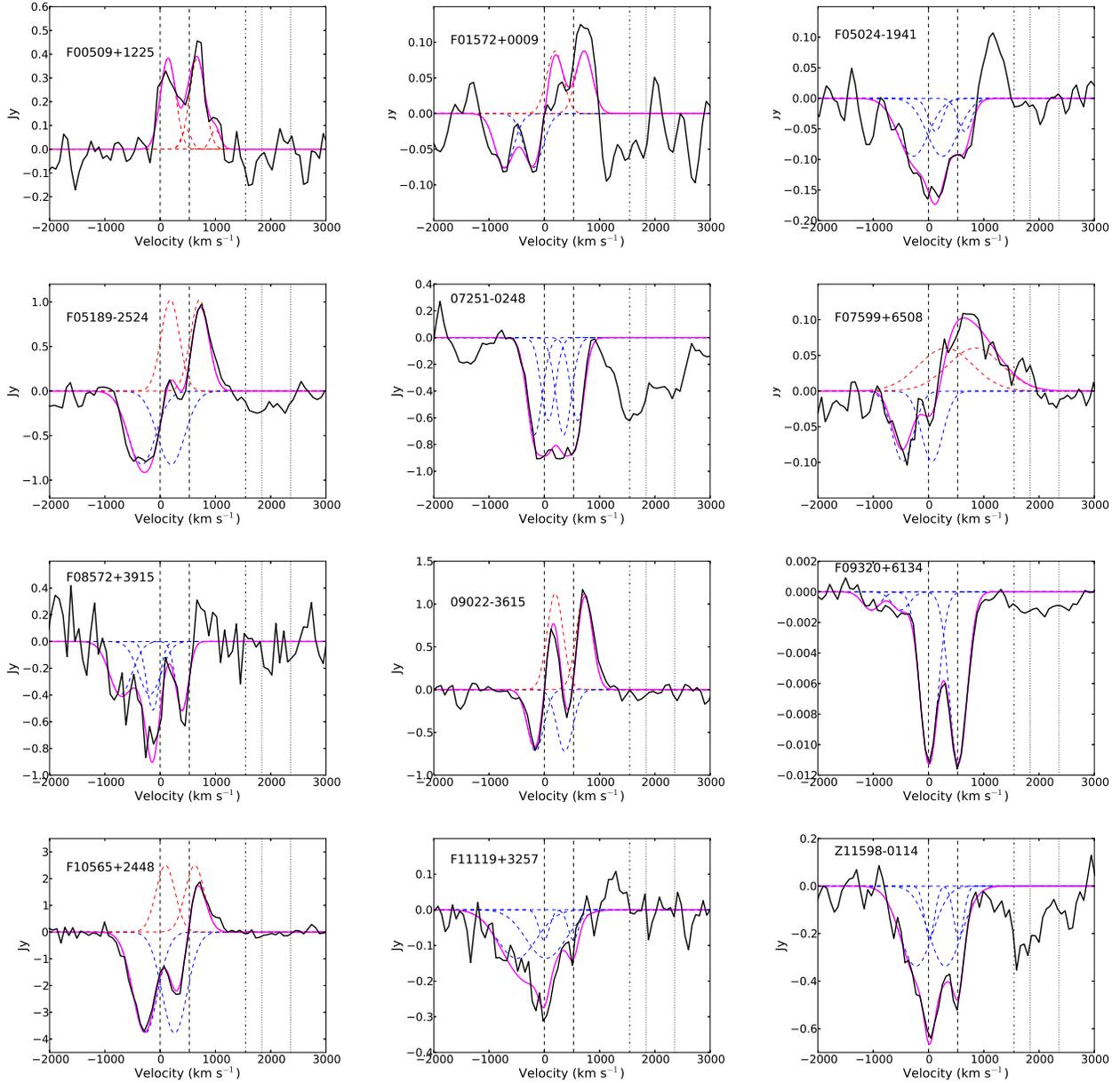}
% \centering
%\includegraphics[width=1.0\textwidth,angle=180]{fig2_1.eps}
\caption{ Spectral fits to the OH 119 $\mu$m profiles in the 43
  objects of our sample. In each panel, the solid black line
  represents the data and the solid purple line is the best
  multi-component Gaussian fit to these data. The blue dash line
  represents the absorption component(s) used in this fit while the
  red dash line represents the emission component(s). The origin of
  the velocity scale corresponds to OH 119.233 $\mu$m at the systemic
  velocity. The two vertical dashed and dotted lines mark the
  positions of the $^{16}$OH and $^{18}$OH doublets, respectively. The
  vertical dot-dash line marks the position of CH$^+$ 119.848
  $\mu$m. }
%\label{fig:tmp}
\end{figure*}

\setcounter{figure}{1}
\begin{figure*}
\epsscale{1.0}
\plotone{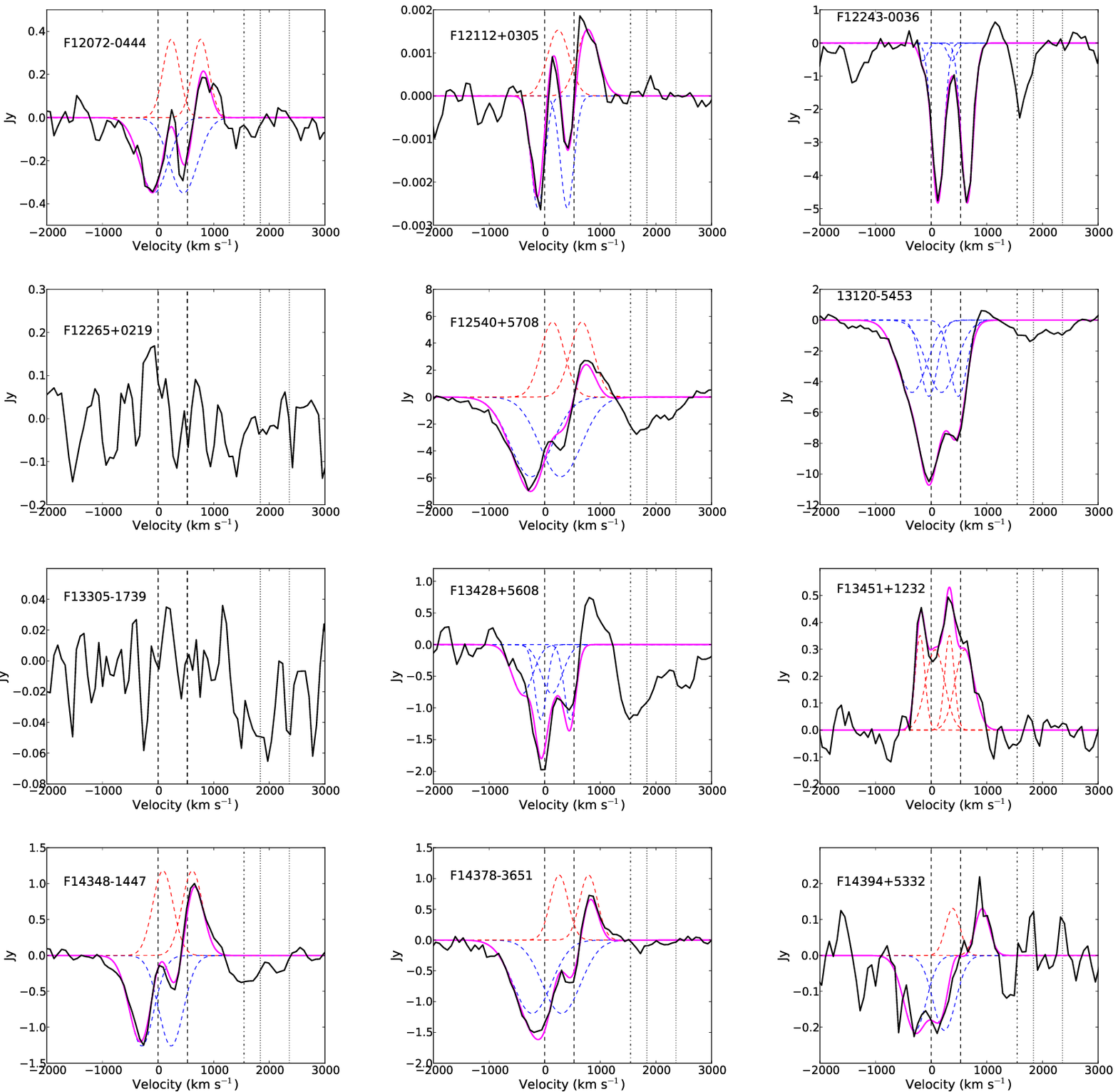}
% \centering
%\includegraphics[width=1.0\textwidth,angle=180]{fig2_2.eps}
\caption{ }
%\label{fig:tmp}
\end{figure*}

\setcounter{figure}{1}
\begin{figure*}
\epsscale{1.0}
\plotone{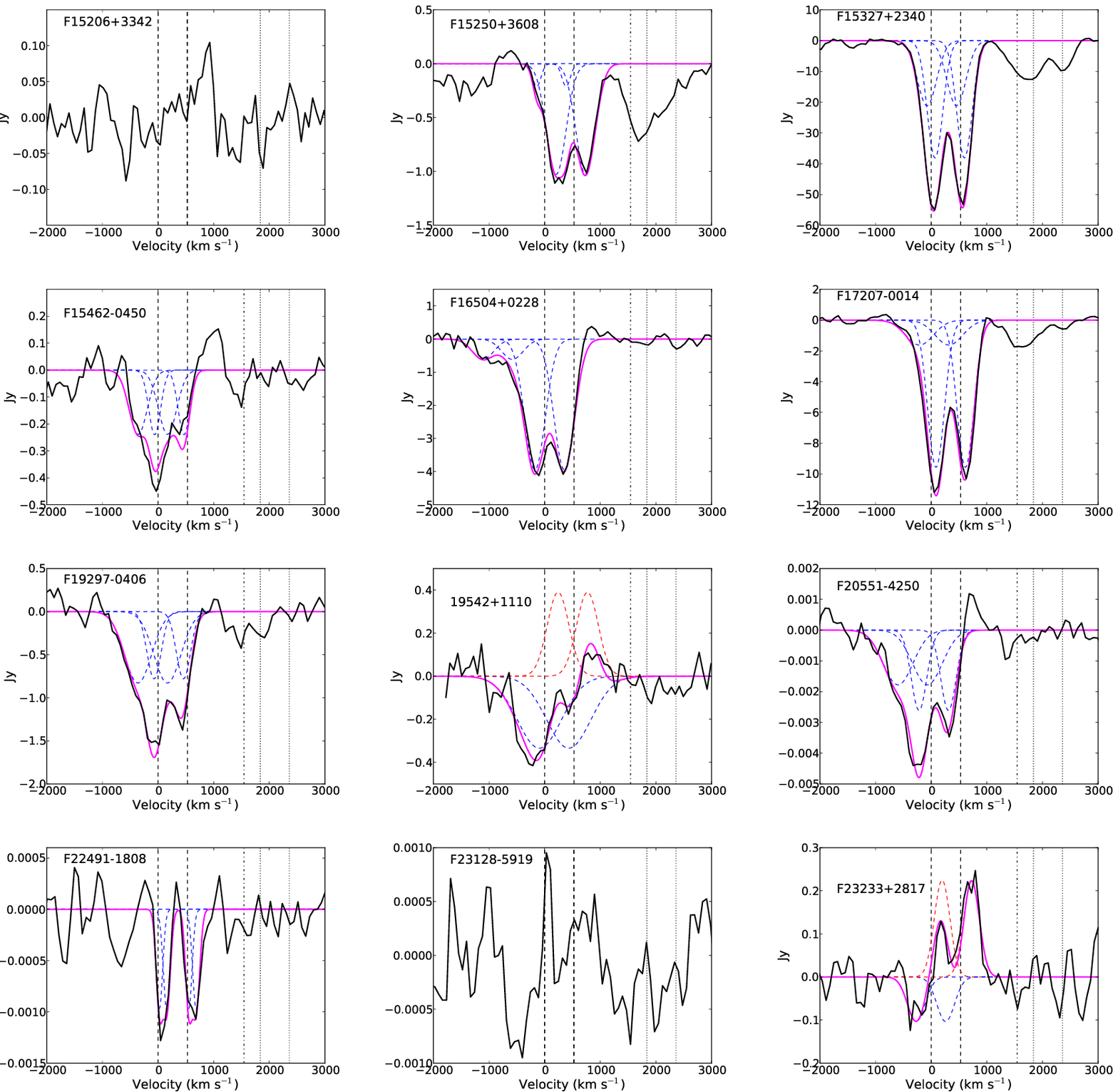}
% \centering
%\includegraphics[width=1.0\textwidth,angle=180]{fig2_3.eps}
\caption{ }
%\label{fig:tmp}
\end{figure*}

\setcounter{figure}{1}
\begin{figure*}
\epsscale{1.0}
\plotone{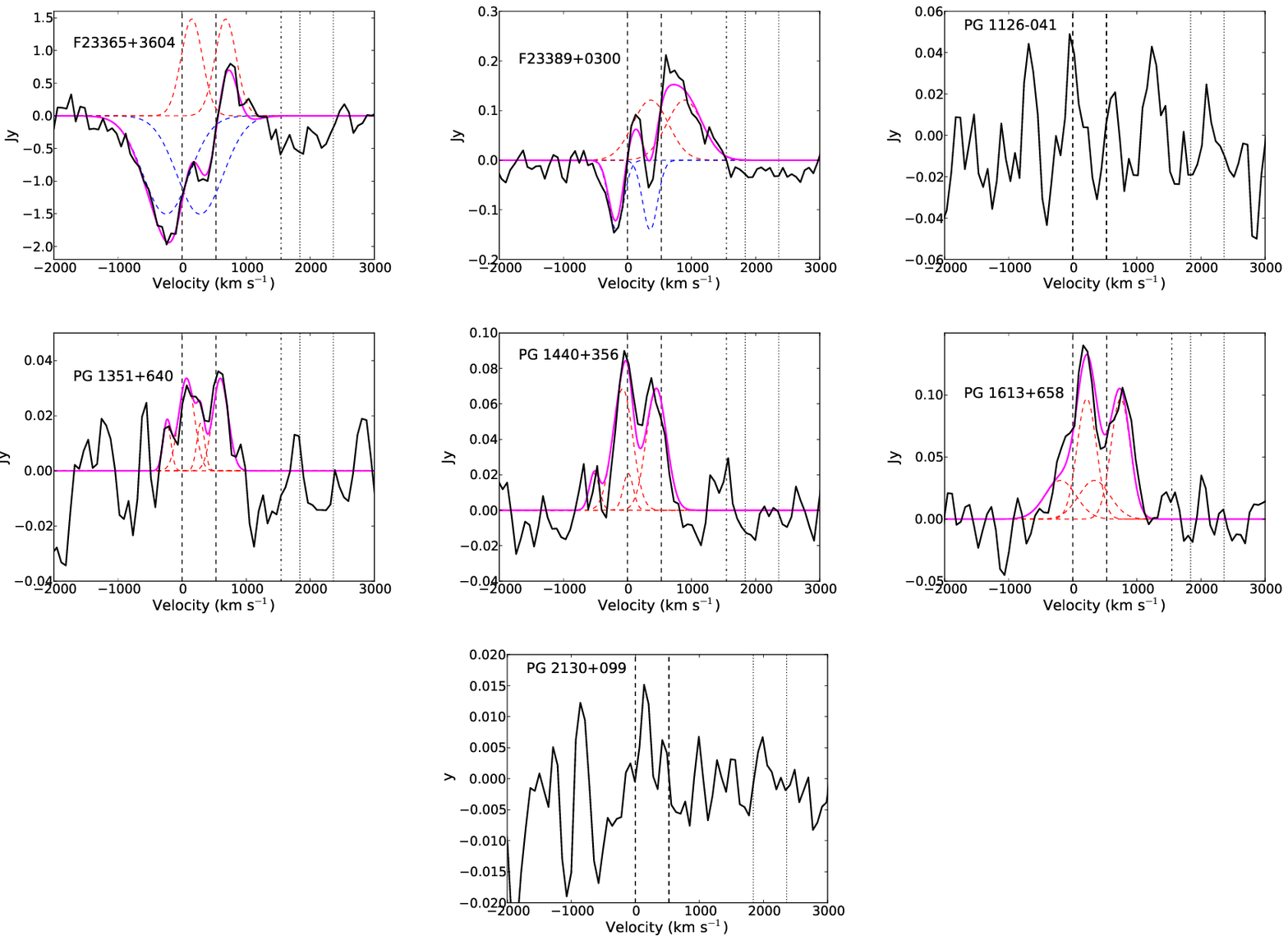}
% \centering
%\includegraphics[width=1.0\textwidth,angle=180]{fig2_4.eps}
\caption{ }
%\label{fig:tmp}
\end{figure*}

\begin{figure*}
\epsscale{1.2}
\plotone{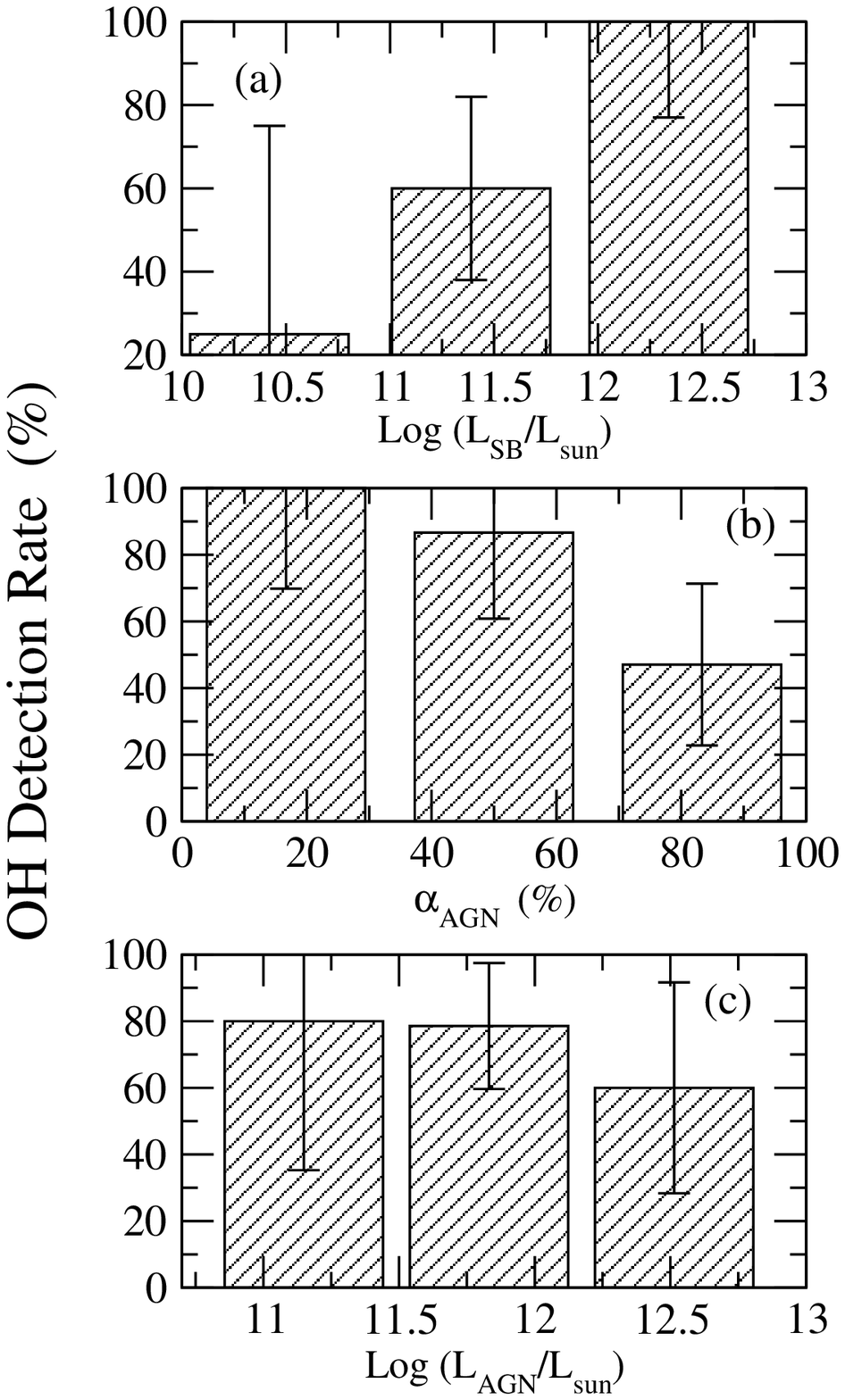}
% \centering
%\includegraphics[width=1.1\textwidth,angle=180]{fig3.eps}
\caption{ Histograms showing the overall detection rates of OH 119
  $\mu$m as a function of ($a$) the starburst luminosities, ($b$) the
  AGN fraction, and ($c$) the AGN luminosities. The uncertainties on
  the detection rates assume a binomial distribution. No significant
  trend is found with any of these parameters (the statistics are poor
  at low starburst luminosities.}
%\label{fig:tmp}
\end{figure*}

\begin{figure*}
\epsscale{1.1}
\plotone{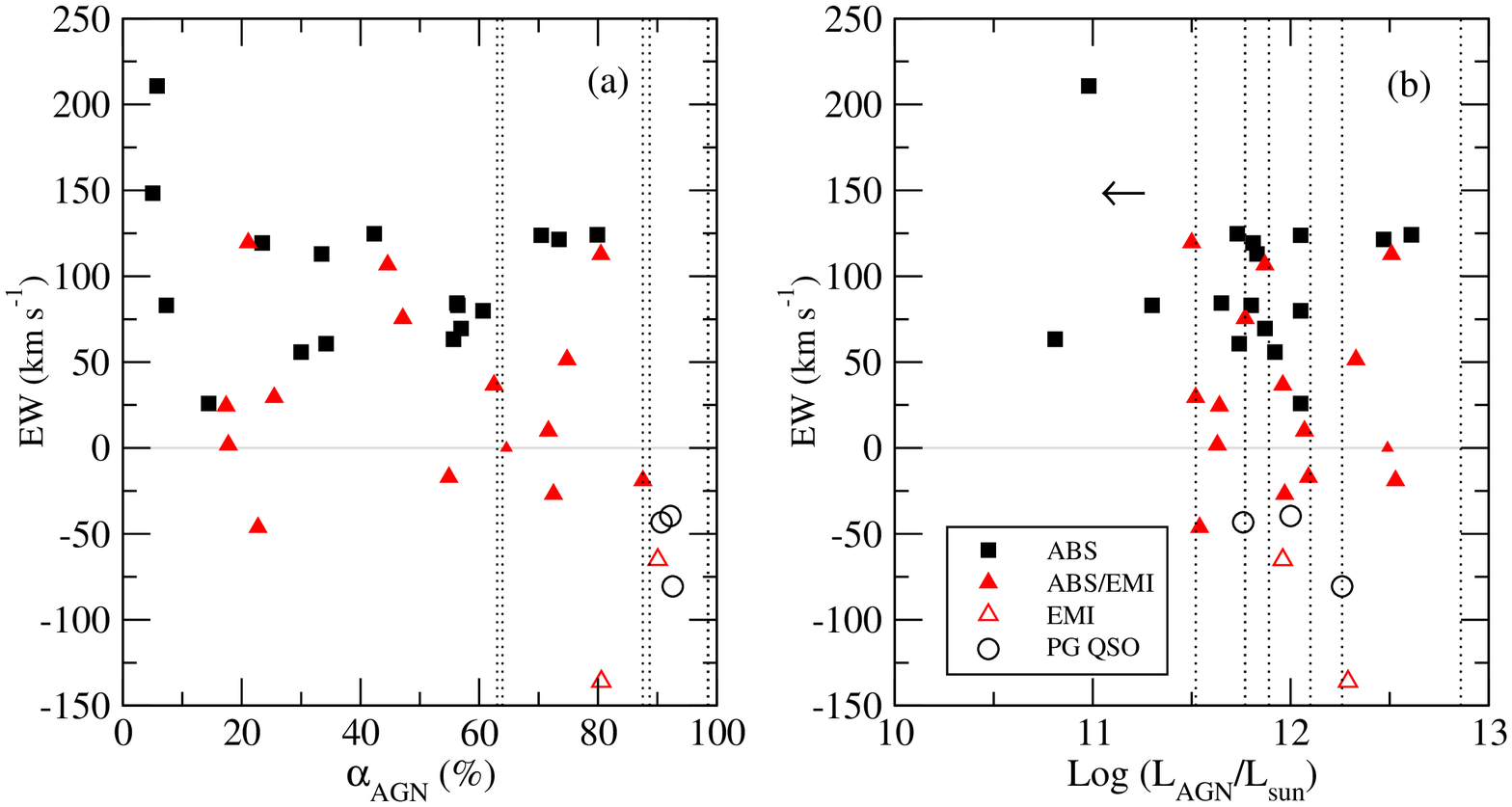}
% \centering
%\includegraphics[width=1.1\textwidth,angle=180]{fig4.eps}
\caption{Total (absorption + emission) equivalent widths of OH 119
  $\mu$m (positive values indicate absorption, negative values
  indicate emission) as a function of ($a$) the AGN fractions and
  ($b$) the AGN luminosities. The meanings of the symbols are as
  follows: Filled black squares, filled red triangles, and open red
  triangles represent ULIRGs with OH 119 $\mu$m seen purely in
  absorption, composite absorption/emission, and purely in emission,
  respectively. The OT2 QSOs, where OH 119 $\mu$m is purely in
  emission, are shown separately as open black circles.  The vertical
  dotted lines indicate the positions of the 6 sources with undetected
  OH 119 $\mu$m (two of these sources have nearly exactly the same AGN
  fraction). The horizontal grey line marks the null OH equivalent
  width.}
%\label{fig:tmp}
\end{figure*}

\begin{figure*}
\epsscale{1.1}
\plotone{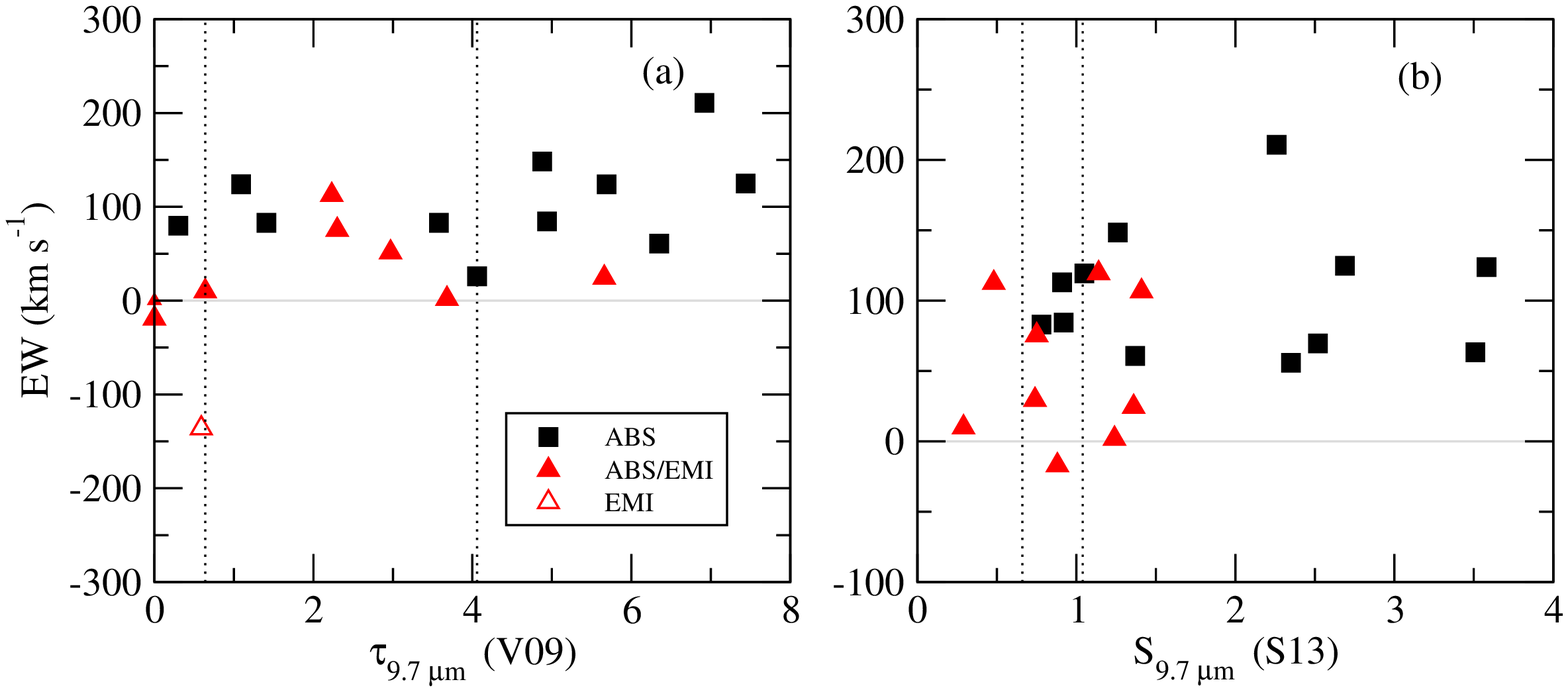}
% \centering
%\includegraphics[width=1.1\textwidth,angle=180]{fig5.eps}
\caption{Total (absorption + emission) equivalent widths of OH 119
  $\mu$m as a function of ($a$) the optical depths of the silicate
  feature at 9.7 $\mu$m measured by V09 (the strength of the silicate
  absorption feature increases to the right) and ($b$) the depths, on
  a logarithmic scale, of the 9.7 $\mu$m features relative to the
  local continuum measured by Stierwalt et al. (2013). In both panels,
  the strength of the silicate absorption feature increases to the
  right (see \S 4.2 for definitions of the silicate-related
  quantities). The sign convention for the equivalent widths and the
  meanings of the symbols are the same as in Figure 4. The vertical
  dotted lines indicate the positions of the two sources with silicate
  measurements but undetected OH 119 $\mu$m. The horizontal grey line
  marks the null OH equivalent width.}
%\label{fig:tmp}
\end{figure*}

\begin{figure*}
\epsscale{1.2}
\plotone{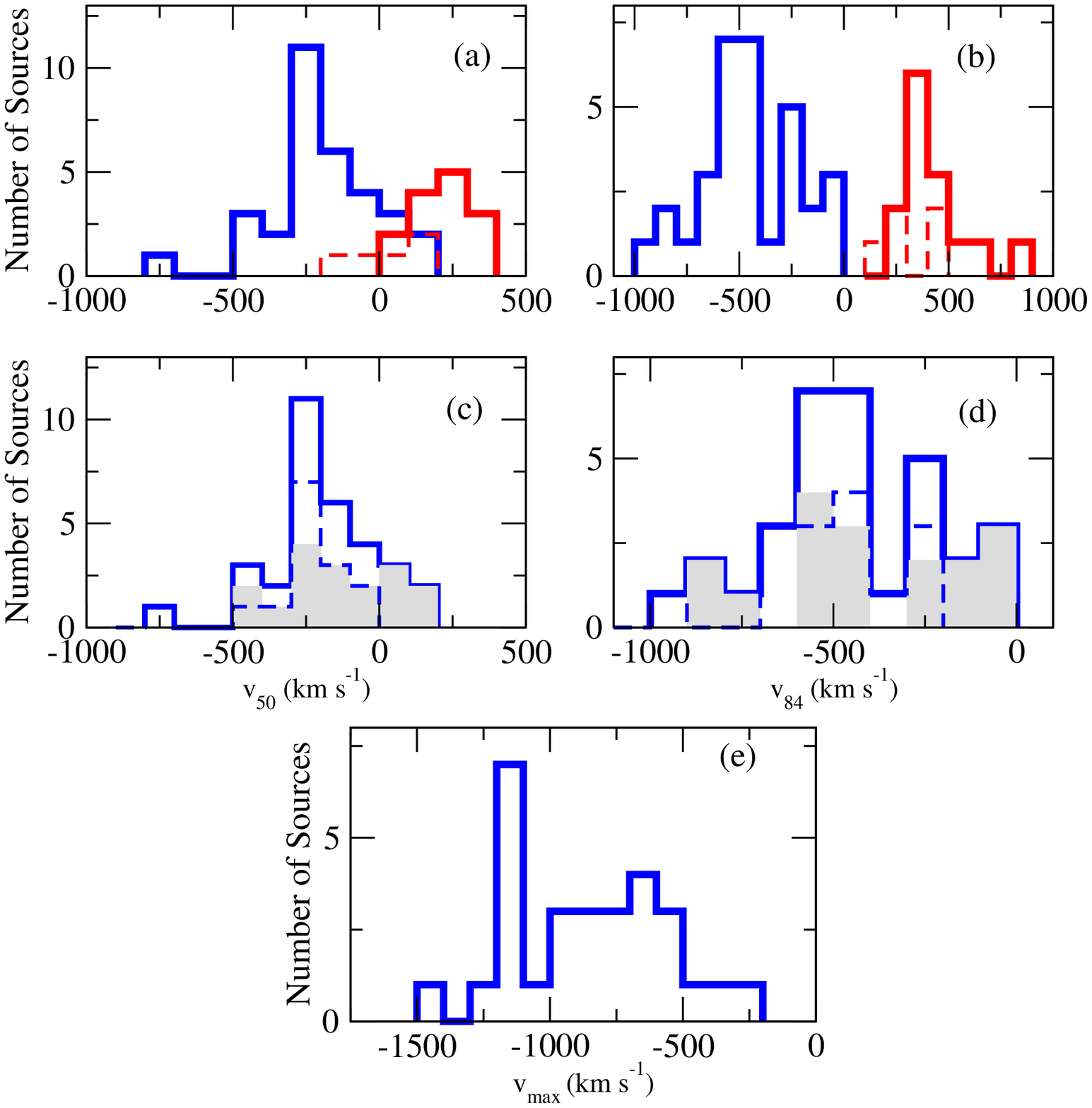}
% \centering
%\includegraphics[width=1.1\textwidth,angle=180]{fig6.eps}
\caption{ Histograms showing ($a$) the distributions of the median
  50\% velocities derived from the multi-Gaussian fits to the OH
  profiles: absorption components (blue) and emission components
  (red), ($b$) same as ($a$) for the 84\% velocities, ($c$) the
  distributions of the median 50\% velocities derived from the
  absorption components in the fits: pure absorption profiles (hatched
  grey) and P~Cygni profiles (dashed blue), ($d$) same as ($c$) for
  the 84\% velociities, and ($e$) the distribution of the terminal
  absorption velocities. Typical uncertainties on $v_{50}$ and
  $v_{84}$ are $\pm$50 km s$^{-1}$ and $\pm$200 km s$^{-1}$ on $v_{\rm
    max}$ (see Table 2). }
%\label{fig:tmp}
\end{figure*}

\begin{figure*}
\epsscale{1.2}
\plotone{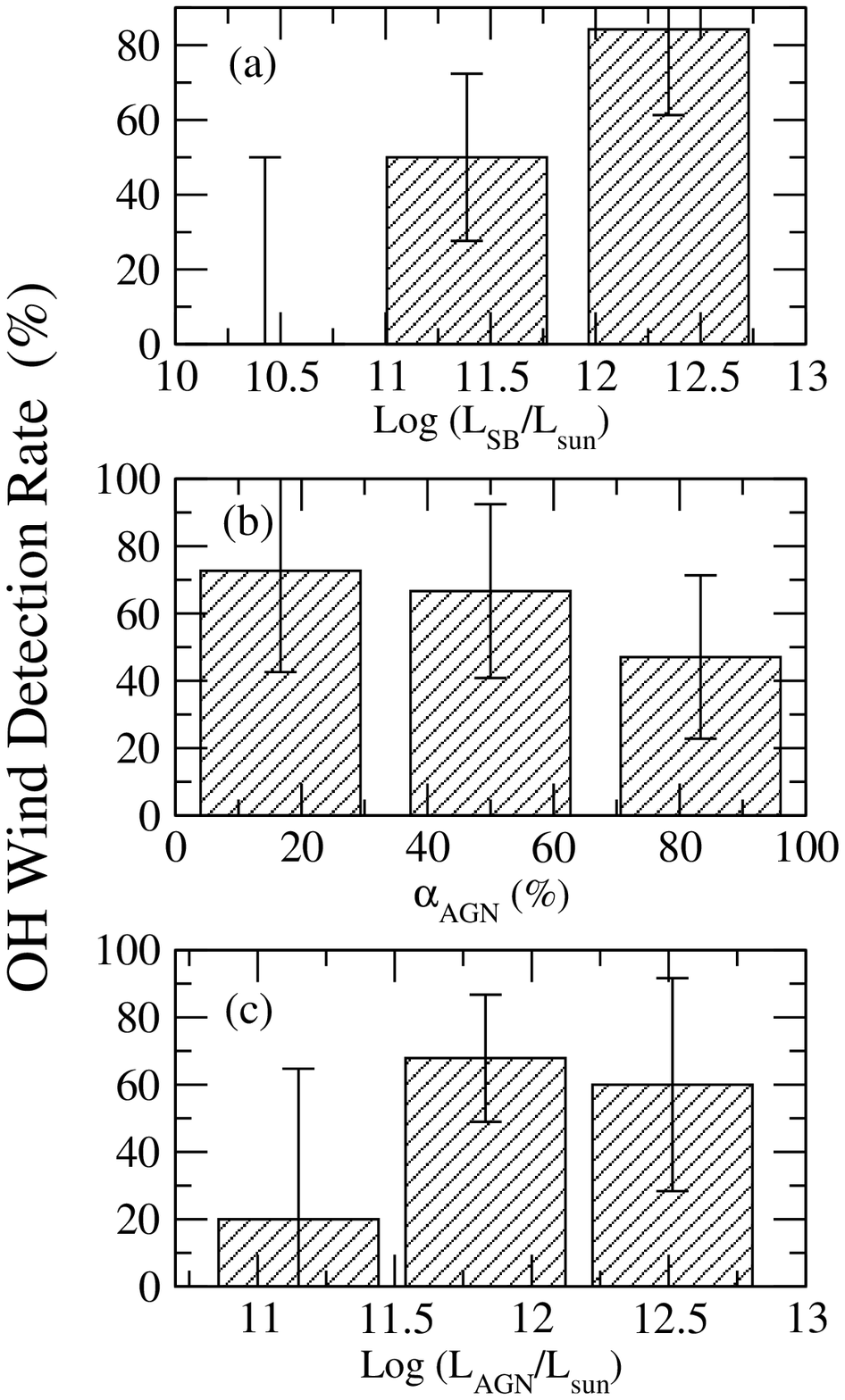}
% \centering
%\includegraphics[width=1.1\textwidth,angle=180]{fig7.eps}
\caption{ Histograms showing the OH wind detection rates, based on the
  presence of an OH 119 $\mu$m absorption feature with a median
  velocity $v_{50}$ more blueshifted than $-$50 km s$^{-1}$, as a
  function of ($a$) the starburst luminosities, ($b$) the AGN
  fraction, and ($c$) the AGN luminosities. The uncertainties on the
  detection rates assume a binomial distribution. No significant trend
  is found with any of these parameters (the statistics are poor at
  low starburst and AGN luminosities). }
 %\label{fig:tmp}
\end{figure*}

\begin{figure*}
\epsscale{1.2}
\plotone{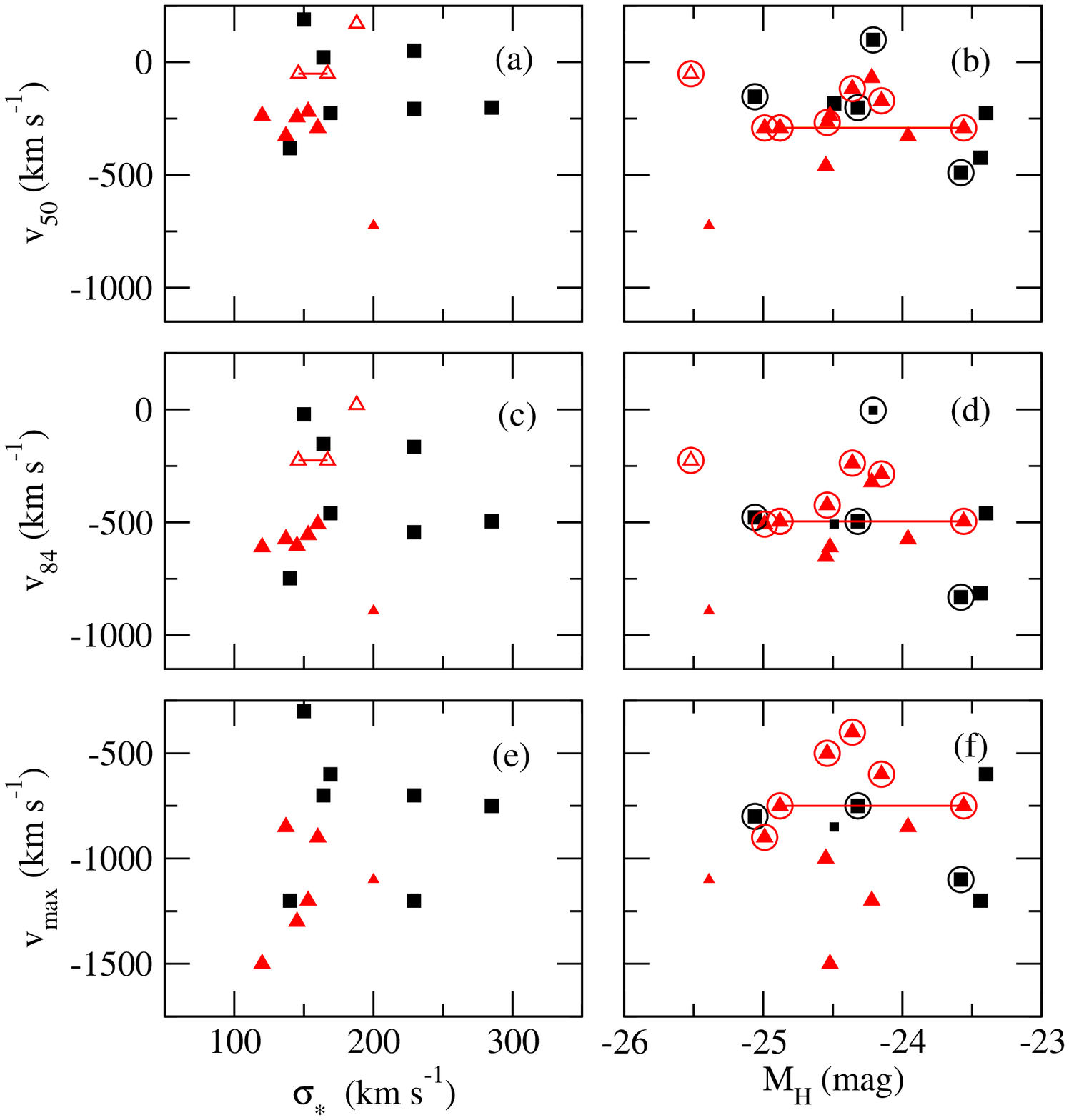}
% \centering
%\includegraphics[width=1.2\textwidth,angle=180]{fig8.eps}
\caption{ The 50\%, 84\%, and terminal OH outflow velocities as a
  function of ($a$)-($c$)-($e$) the near-infrared derived stellar
  velocity dispersions measured by Dasyra et al. (2005a, 2005b, and
  2006; the typical uncertainties on the velocity dispersions are
  $\sim$50 km s$^{-1}$) and ($b$)-($d$)-($f$) the H-band absolute
  magnitudes of the host galaxies from Veilleux et al. (2006) or from
  Veilleux et al. (2002) assuming H -- K' = 0.35 mag (the typical
  uncertainties on these magnitudes are 0.1 and 0.3 mag.,
  respectively). The meanings of the symbols are the same as in Figure
  4. In the lower panels, data points with red circles around them
  indicate values based on Veilleux et al. (2002). The two data points
  joined by a segment in the left (right) panels correspond to
  F13451$+$1232 E and W (F14394$+$5332 E and W). Typical uncertainties
  on $v_{50}$ and $v_{84}$ are $\pm$50 km s$^{-1}$ and $\pm$200 km
  s$^{-1}$ on $v_{\rm max}$. The smaller symbols have larger
    uncertainties (values followed by double colons in Table 2).}
%\label{fig:tmp}
\end{figure*}

\begin{figure*}
\epsscale{1.5}
\plotone{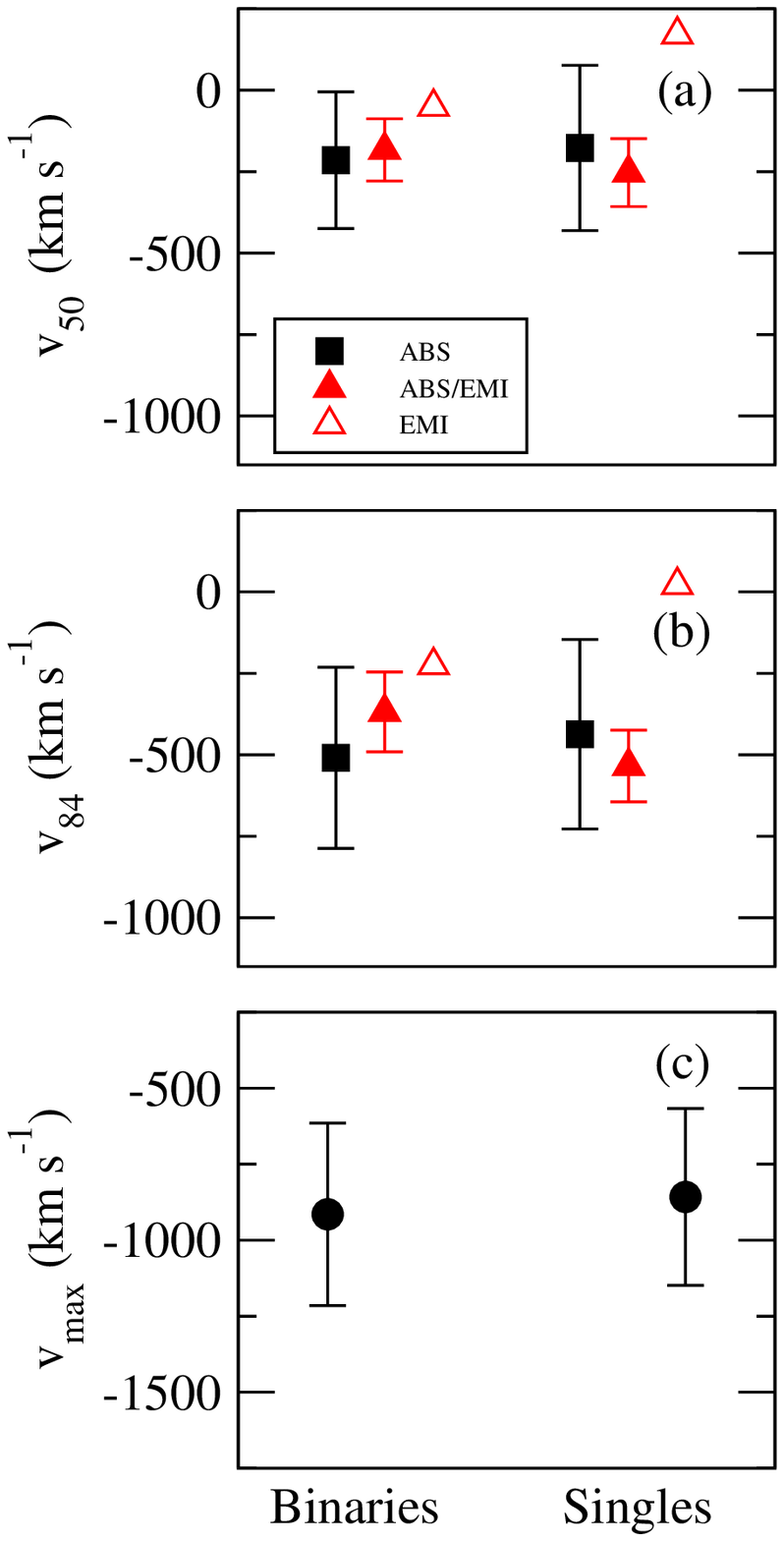}
% \centering
%\includegraphics[width=1.4\textwidth,angle=180]{fig9.eps}
\caption{ The 50\%, 84\%, and terminal OH outflow velocities as a
  function of interaction class: binaries or single sources. The
  meanings of the symbols are the same as in Figure 4. The error bars
  indicate the 1-sigma scatter in each distribution. There are only
  two ULIRGs with OH purely in emission: one of them (F13451$+$1232)
  is a binary while the other has a single nucleus (F00509$+$1225 =
  I~Zw 1). Typical uncertainties on $v_{50}$ and $v_{84}$ are $\pm$50
  km s$^{-1}$ and $\pm$200 km s$^{-1}$ on $v_{\rm max}$ (see
    Table 2). }
%\label{fig:tmp}
\end{figure*}

\begin{figure*}
\epsscale{1.3}
\plotone{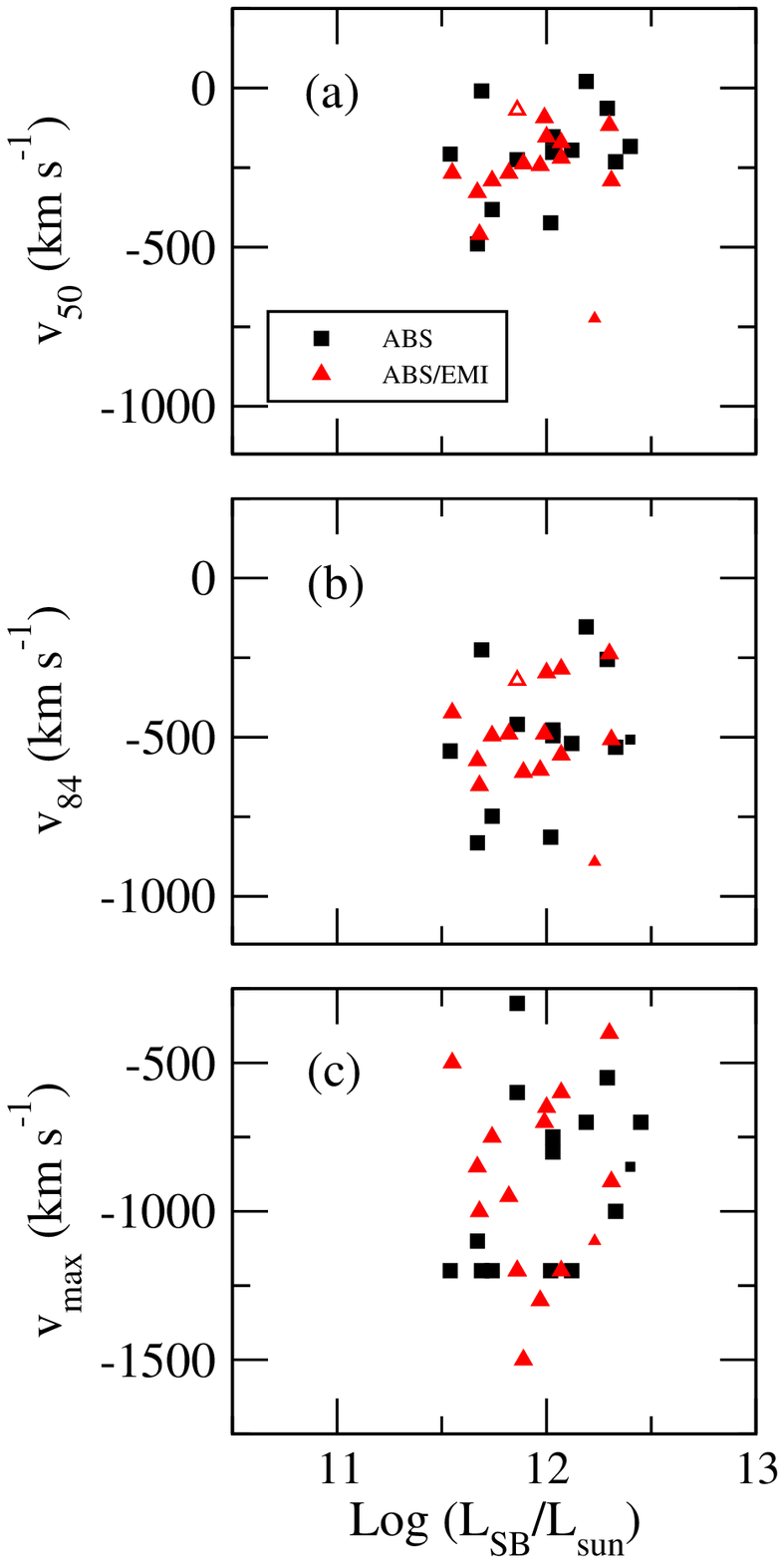}
% \centering
%\includegraphics[width=1.4\textwidth,angle=180]{fig10.eps}
\caption{ The 50\%, 84\%, and terminal OH outflow velocities as a
  function of the starburst luminosities. The meanings of the symbols
  are the same as in Figure 4.  The four objects with significant
    inflows (50\% OH velocities above 50 km s$^{-1}$) are not shown
    here. None of these quantities is significantly correlated with
    the starburst luminosities (P[null] $>$ 0.05; see Table 3). 
  F12072$-$0444, indicated by an open red triangle, was not included
  in the evaluations of these correlations because it is not clear
  which of the two nuclei is responsible for the OH absorption feature
  (both nuclei are included in the PACS entrance aperture). Typical
  uncertainties on $v_{50}$ and $v_{84}$ are $\pm$50 km s$^{-1}$ and
  $\pm$200 km s$^{-1}$ on $v_{\rm max}$. The smaller symbols have
    larger uncertainties (values followed by double colons in Table
    2). }
%\label{fig:tmp}
\end{figure*}

\begin{figure*}
\epsscale{1.3}
\plotone{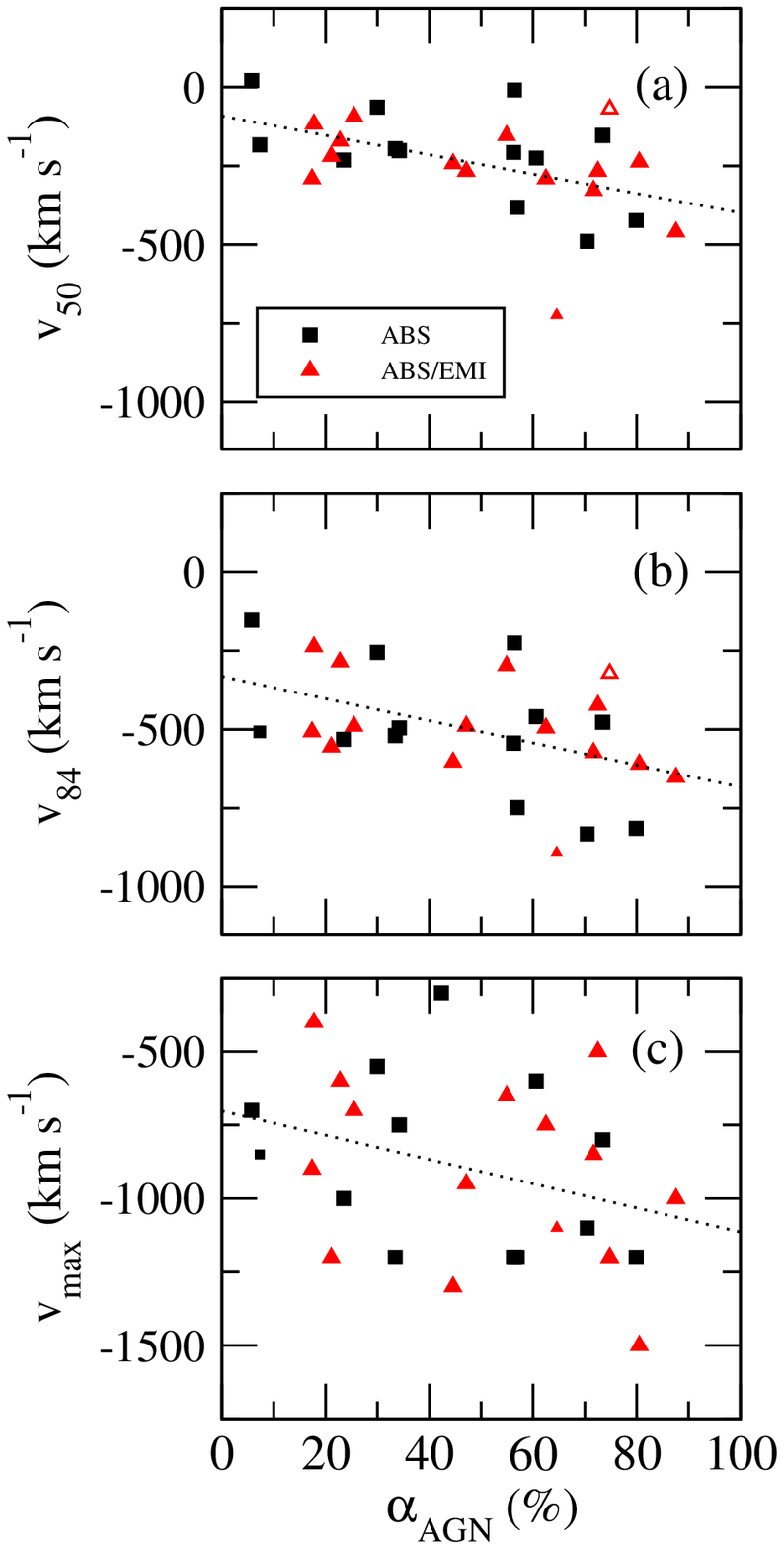}
% \centering
%\includegraphics[width=1.4\textwidth,angle=180]{fig11.eps}
\caption{ The 50\%, 84\%, and terminal OH outflow velocities as a
  function of the AGN fractions. The meanings of the symbols are the
  same as in Figure 10. The four objects with significant inflows
  (50\% OH velocities above 50 km s$^{-1}$) are not shown here. Dotted
  lines indicate the best linear fits through the data. A significant
  linear correlation is present with $v_{50}$ and $v_{84}$ (P[null]
  $\le$ 0.05), but only tentatively with $v_{\rm max}$ (P[null] $\sim$
  0.07; Table 3). The coefficients of the correlations are listed in
  Table 4.  F12072$-$0444, indicated by an open red triangle, was not
  included in the evaluations of these correlations because it is not
  clear which of the two nuclei is responsible for the OH absorption
  feature (both nuclei are included in the PACS entrance
  aperture). Typical uncertainties on $v_{50}$ and $v_{84}$ are
  $\pm$50 km s$^{-1}$ and $\pm$200 km s$^{-1}$ on $v_{\rm max}$. The
  smaller symbols have larger uncertainties (values followed by double
  colons in Table 2). }
%\label{fig:tmp}
\end{figure*}

\begin{figure*}
\epsscale{1.3}
\plotone{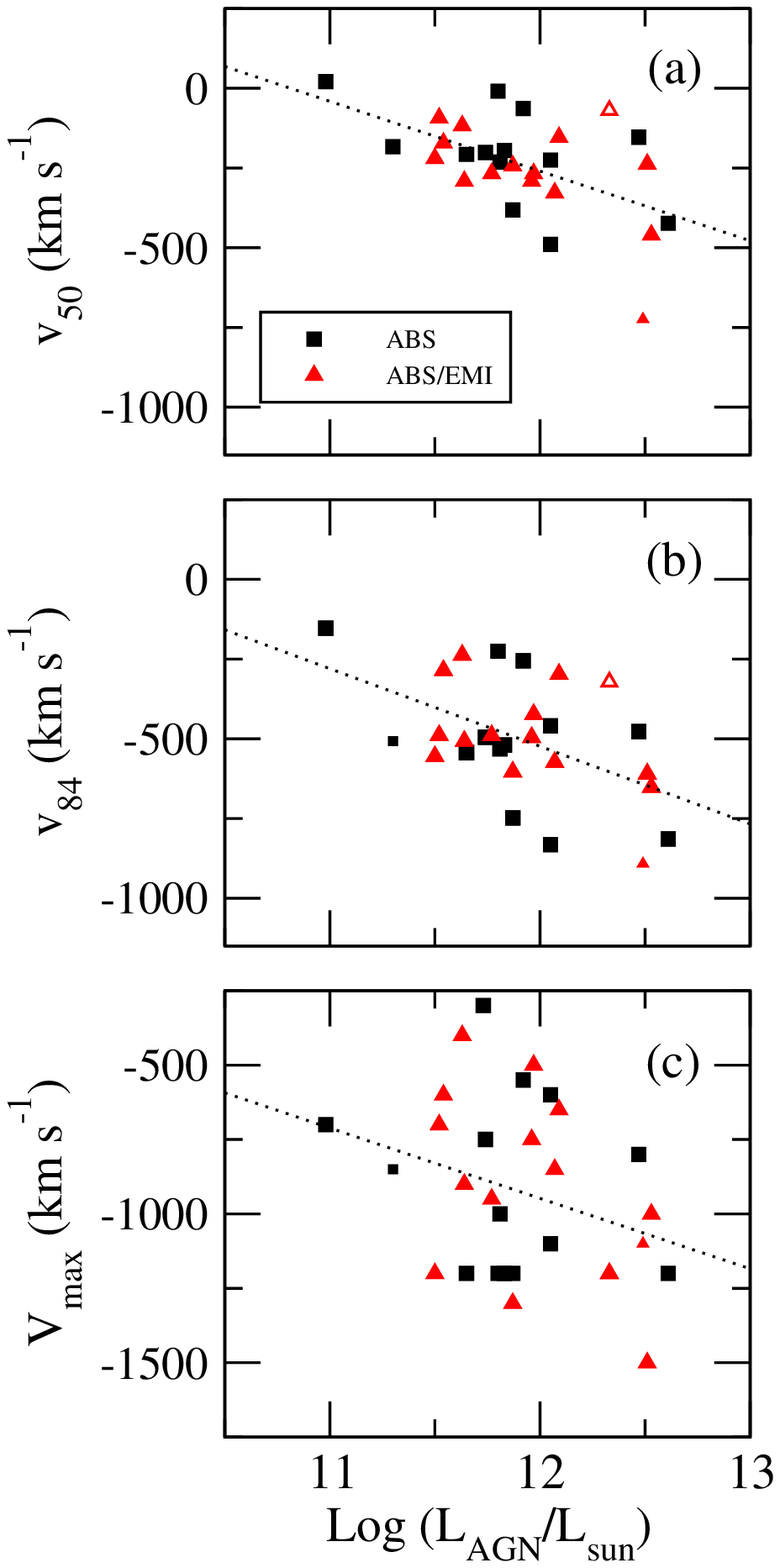}
% \centering
%\includegraphics[width=1.4\textwidth,angle=180]{fig12.eps}
\caption{ The 50\%, 84\%, and terminal OH outflow velocities as a
  function of the AGN luminosities. The meanings of the symbols are
  the same as in Figure 10. The four objects with significant inflows
  (50\% OH velocities above 50 km s$^{-1}$) are not shown here. Dotted
  lines indicate the best linear fits through the data. A significant
  linear correlation is present with $v_{50}$ and $v_{84}$ (P[null]
  $\le$ 0.05), but only tentatively with $v_{\rm max}$ (P[null] $\sim$
  0.09; Table 3). The coefficients of the correlations are listed in
  Table 4.  F12072$-$0444, indicated by an open red triangle, was not
  included in the evaluations of these correlations because it is not
  clear which of the two nuclei is responsible for the OH absorption
  feature (both nuclei are included in the PACS entrance
  aperture). Typical uncertainties on $v_{50}$ and $v_{84}$ are
  $\pm$50 km s$^{-1}$ and $\pm$200 km s$^{-1}$ on $v_{\rm max}$. The
  smaller symbols have larger uncertainties (values followed by double
  colons in Table 2). }
%\label{fig:tmp}
\end{figure*}

\begin{figure*}
\epsscale{1.1}
\plotone{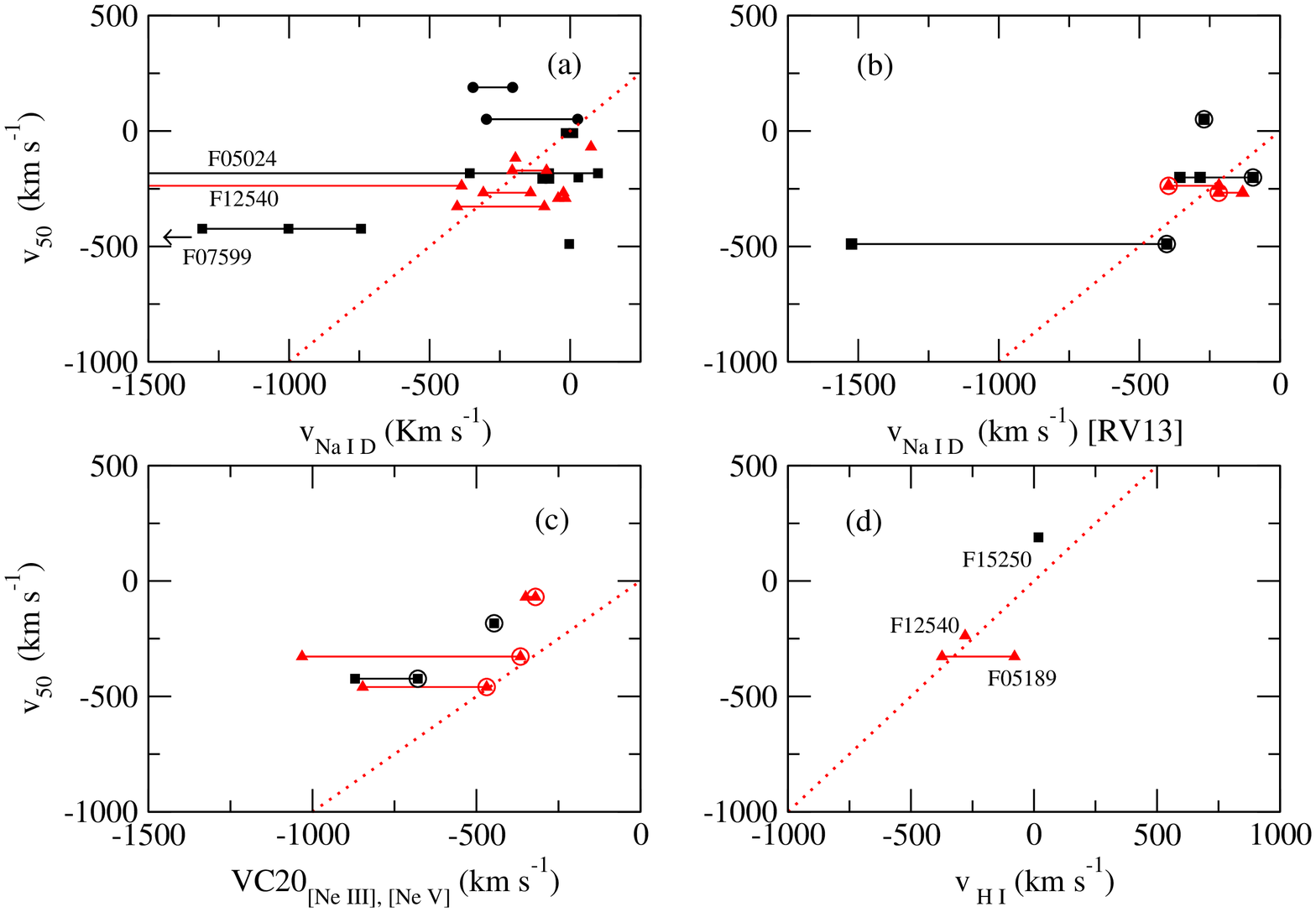}
% \centering
%\includegraphics[width=1.1\textwidth,angle=180]{fig13.eps}
\caption{ Multi-phase comparison of the outflow velocities.  The 50\%
  OH outflow velocities are compared with ($a$) the 50\% Na I D
  outflow velocities measured by Rupke et al. (2005a, 2005b, 2005c)
  and Krug et al. (2013, in prep.) from integrated spectra (in
  F05024$-$1941, F07599$+$6508, and 12540$+$5708, there are multiple
  Na I components seen at diffferent velocities), ($b$) the
  spatially-averaged 50\% Na I D (encircled) and H$\alpha$ outflow
  velocities measured by RV13 outside of the nucleus using the IFU
  data, ($c$) the outflow velocities measured by Spoon \& Holt (2009)
  from the emission profiles of [Ne~III] 15.5 $\mu$m (encircled) and
  [Ne~V] 14.3 $\mu$m in integrated {\em Spitzer} spectra, and ($d$)
  the H~I 21-cm outflow velocities measured by Teng, Veilleux, \&
  Baker (2013) from GBT data.  The meanings of the symbols are the
  same as in Figure 4. The red dotted line is the 1:1 line. }
%\label{fig:tmp}
\end{figure*}

\clearpage


\clearpage

\begin{references}
\refpar
Aalto, S, et al. 2012, A\&A, 537, 44
% \refpar
% Allen, J. T., et al. 2011, MNRAS, 410, 860
% \refpar
% Bautista, M., et al. 2010, ApJ, 713, 25
% \refpar
% Boksenberg, A., et al. 1977, MNRAS, 178, 451
% \refpar
% Boroson, T. A., et al. 1991, ApJ, 370, L19
% \refpar
% Braito, V., et al. 2004, A\&A, 420, 79
% \refpar
% Carilli, C. L., Wrobel, J. M., \& Ulvestad, J. S. 1998, AJ, 115, 928
% \refpar
% Chelouche, D., \& Netzer, H. 2003, MNRAS, 344, 233
\refpar
Chen, Y.-M., et al. 2010, ApJ, 140, 445
\refpar
Chevalier, R. A., \& Clegg, A. W. 1985, Nature, 317, 44
\refpar
Cicone, C., et al. 2012, A\&A, 543, 99
\refpar
Cicone, C., et al. 2013, preprint
\refpar
Combes, F., et al. 2013, A\&A, 550, 41
% \refpar
% Conroy, C., et al. 2010, ApJ, 718, 184
% \refpar
% Cooper, J. L., et al.\  2009, ApJ, 703, 330
% \refpar
% Crenshaw, M., Kraemer, S. B., \& George, I. M. 2003, ARAA, 41, 117
\refpar
Daddi, E., et al. 2010, ApJ, 713, 686
\refpar
Dasyra, K. M., et al. 2006a, ApJ, 638, 745
\refpar
Dasyra, K. M., et al. 2006b, ApJ, 651, 835
\refpar
Dasyra, K. M., et al. 2007, ApJ, 657, 102
% \refpar
% Davies, R. I., Tacconi, L. J., \& Genzel, R. 2004, ApJ, 613, 781
% \refpar
% Davis, S. W., Woo, J.-H., \& Blaes, O. M. 2007, ApJ, 668, 682
% \refpar
% Dermer, C. D., et al. 1997, ApJ, 484, L121
\refpar
De Young, D., \& Heckman, T. M. 1994, ApJ, 431, 598
% \refpar
% Di Matteo, T., Springel, V., \& Hernquist, L. 2005, Nature, 433, 604
\refpar
Downes, D., \& Solomon, P. M. 1998, ApJ, 507, 615
% \refpar
% Eales, S. A., \& Arnaud, K. A. 1988, ApJ, 324, 193
% \refpar
% Edmonds, D., et al. 2011, ApJ, 739, 7
% \refpar
% Elvis, M. 2000, ApJ, 545, 63
\refpar
Erb, D. K., et al. 2012, ApJ, 759, 26
% \refpar
% {\bf Evans, A. S., et al. 2001, AJ, 121, 3285}
\refpar
Fabian, A. C. 1999, MNRAS, 308, L39
% \refpar
% Fall, S. M., Pei, Y. C., \& McMahon, R. G. 1989, ApJ, 341, L5
\refpar
Faucher-Gigu\`ere, C.-A., \& Quataert, E. 2012, MNRAS, 425, 605
\refpar
Faucher-Gigu\`ere, C.-A., Quataert, E., \& Murray, N. 2012, MNRAS, 420, 1347
\refpar
Feruglio, C., et al.\ 2010,  A\&A,  518, L155
\refpar
Fischer, J., et al.\ 2010,  A\&A,  518, L41 (F10)
% \refpar
% Forster, K., Rich, R. M., \& McCarthy, J. K. 1995, ApJ, 450, 74
% \refpar
% Gallagher, S. C., et al. 2002, ApJ, 569, 655
% \refpar
% Gallagher, S. C., et al. 2005, ApJ, 633, 71
% \refpar
% Garmire, G. P., et al. 2003, SPIE, 4851, 28
\refpar
Ginsburg, A., \& Mirocha, J. 2011, Astrophysics Source Code Library, record ascl:1109.001
% \refpar
% Gibson, R. R., et al. 2008, ApJ, 675, 985
% \refpar
% Gibson, R. R., et al. 2009, ApJ, 692, 758
% \refpar
% Goodrich, R. W., \& Miller, J. S. 1994, ApJ, 434, 82
\refpar
Gonzalez-Alfonso, E., et al. 2012, A\&A, 541, 4
\refpar
Gonzalez-Alfonso et al. 2013, A\&A, submitted, refereed
% \refpar
% Green, J. C., et al. 2012, ApJ, 744, 60
% \refpar
% Grimes, J. P., et al. 2005, ApJ, 628, 187
\refpar
Hailey-Dunsheath, S., et al. 2012, ApJ, 755, 57
% \refpar
% Hall, P. B., et al. 2002, ApJS, 141, 267
% \refpar
% Hall, P. B., et al. 2011, MNRAS, 411, 2653
% \refpar
% Hamann, F., Korista, K. T., \& Morris, S. L. 1993, ApJ, 415, 541
% \refpar
% Hamann, F., \& Sabra, B. 2004, in AGN Physics with the Sloan Digital
% Sky Survey, ASP Conf. Series, 311, p. 203
% \refpar
% Hamilton, D., \& Keel, W. C. 1987, ApJ, 321, 211
\refpar
Heckman, T. 2002, ASP Conf. proceedings vol. 254, 292 
\refpar
Hopkins, P. F., et al. 2009, MNRAS, 398, 303
% \refpar
% Hutchings, J. B., \& Neff, S. G. 1987, AJ, 92, 14
% \refpar
% Iwasawa, K., et al. 2011, A\&A, 528, 137
% \refpar
% Kashyap, V., et al. 2011, Chandra Interactive Analysis of Observations
% (CIAO), Sec. 4.3
\refpar
Kauffmann, G., et al. 2003, MNRAS, 341, 54
% \refpar
% Kim, D.-W. 2012, Astrophysics and Space Science Library, 378, 121 
\refpar
King, A. 2003, ApJ, 596, L27
% \refpar
% Klockner, H.-R., Baan, W. A., \& Garrett, M. A. 2003, Nature, 421, 821
% \refpar
% Kollatschny, W., Dietrich, M., \& Hagen, H. 1992, A\&A, 264, L5
\refpar
Kornei, K. A., et al. 2012, ApJ, 758, 135
% \refpar
% Kramer, R. H., \& Haiman, Z. 2009, MNRAS, 400, 1493
% \refpar
% Krolik, J. H., \& McKee, C. F. 1978, ApJS, 37, 459
% \refpar
% Kruczek, N. E., et al. 2011, AJ, 142, 130
\refpar
Krug, H. B., Rupke, D. S. N., \& Veilleux, S. 2010,  ApJ,  708, 1145
% \refpar
% Laor, A., et al. 1997, ApJ, 489, 656
\refpar
Laurent, O., et al. 2000, A\&A, 359, 887
\refpar
Law, D. R., et al. 2012, ApJ, 745, 85
% \refpar
% Leighly, K., et al. 2007a, ApJ, 663, 103
% \refpar
% Leighly, K., et al. 2007b, ApJS, 173, 1
% \refpar
% Leitherer, C., et al. 2001, ApJ, 550, 724
\refpar
Leroy, A. K., et al. 2008, AJ, 136, 2782
% \refpar
% Li, J., et al. 2004, ApJ, 610, 1204
% \refpar
% Lipari, S., Colina, L., \& Macchetto, F. 1994, ApJ, 427, 174
% \refpar
% Lipari, S., et al. 2009, MNRAS, 392, 1295
% \refpar
% Maloney, P. R., \& Reynolds, C. S. 2000, ApJ, 545, L23
% \refpar
% Marcolini, A., et al.\ 2005, MNRAS, 362, 626
\refpar
Markwardt, C. B. 2009, ADASS XVIII, ASP Conf. Ser., 411, 251
\refpar
Martin, C. L. 2005 ApJ, 621, 227
\refpar
Martin, C. L. 2006 ApJ, 647, 222
\refpar
Martin, C. L., et al. 2012, ApJ, 760, 127
% \refpar
% Morabito, L. K., et al. 2011, ApJ, 737, 46
% \refpar
% Morton, D. C. 2003, ApJS, 149, 205
% \refpar
% Murphy, K. D., \& Yaqoob, T. 2009, MNRAS, 397, 1549
% \refpar
% Murray, N., \& Chiang, J. 1995, ApJ, 454, L105
\refpar
Murray, N., Quataert, E., \& Thompson, T. A. 2005,  ApJ,  618, 569
% \refpar
% Murray, N., et al. 1995, ApJ, 451, 498
\refpar
Narayanan, D., et al. 2008, ApJS, 176, 331
\refpar
Netzer, H., et al. 2007, ApJ, 666, 806 
\refpar
Newman, S., et al. 2012, ApJ, 761, 43
% \refpar
% Piconcelli, E., et al.\ 2005, A\&A, 432, 15
% \refpar
% Piconcelli, E., et al.\ 2013, MNRAS, 428, 1185
 \refpar
Pilbratt, G. L. et al. 2010, A\&A, 518, L1
 \refpar
Poglitsch, A., et al. 2010, A\&A, 518, L2
% \refpar
% Proga, D., \& Kallman 2004, ApJ, 616, 688
% \refpar
% Proga, D. 2007, ApJ, 661, 693
% \refpar
% Proga, D., Ostriker, J. P., \& Kurosawa, R. 2008, ApJ, 676, 101
% \refpar
% Proga, D., Stone, J. M., \& Kallman, T. R. 2000, ApJ, 543, 686
% \refpar
% Ptak, A., et al. 2003, ApJ, 592, 782
% \refpar
% Reichard, T. A., et al. 2003, AJ, 126, 2594
% \refpar
% Richards, G. T., et al. 2002, AJ, 124, 1
% \refpar
% Richards, G. T., et al. 2011, ApJ, 141, 167
% \refpar
% Richards, G. T. 2012, in ``AGN Winds in Charleston'', preprint (arXiv:1201.2595)
% \refpar
% Robert, C., et al. 2003, ApJS, 144, 21
% \refpar
% Rogerson, J. A., et al. 2011, New Astronomy, 16, 128
\refpar
Rothberg \& Fischer 2010, ApJ, 712, 318
\refpar
Rothberg, B., et al. 2013, ApJ, 767, 72
\refpar
Rubin, K. H. R., et al. 2010, ApJ, 719, 1503
\refpar
Rupke, D. S. N., \& Veilleux, S. 2011,  ApJ,  729, L27 (RV11)
\refpar
Rupke, D. S. N., \& Veilleux, S. 2013a,  ApJ,  768, 75 (RV13)
\refpar
Rupke, D. S. N., \& Veilleux, S. 2013b,  ApJ,  NNN, LNN (arXiv:NNNN.NNNN)
\refpar 
Rupke, D.~S., Veilleux, S., \& Sanders, D.~B.  2002, ApJ,  570, 588
\refpar
Rupke, D.~S., Veilleux, S., \& Sanders, D.~B. 2005a, ApJS, 160, 87
\refpar
Rupke, D.~S., Veilleux, S., \& Sanders, D.~B. 2005b, ApJS, 160, 115
\refpar 
Rupke, D.~S., Veilleux, S., \& Sanders, D.~B.  2005c, ApJ, 632, 751
\refpar
Saintonge, A., et al. 2011, MNRAS, 415, 61
\refpar
Sanders, D. B., \& Mirabel, I. F. 1996, ARAA, 34, 749
\refpar
Sanders, D. B., et al. 1988, ApJ, 325, 74
\refpar
Schwartz, C., \& Martin, C. L. 2004, ApJ, 610, 201
\refpar
Schweitzer, M., et al. 2008, ApJ, 679, 101
\refpar
Scoville, N. Z., et al. 2003, ApJ, 585, L105
\refpar
Spinoglio, L., et al. 2005, ApJ, 623, 123
\refpar
Spoon, H., \& Holt, J. 2009, ApJ, 702, L42
\refpar
Steidel, C. C., et al. 2010, ApJ, 717, 289
\refpar
Strauss, M. et al. 1992, ApJS, 83, 29
\refpar
Strel'nitskii, V. S., \& Sunyaev, R. A. 1973, Soviet Astronomy, 16, 579
% \refpar
% Strickland, D. K., \& Heckman, T. M. 2007, ApJ, 658, 258
%\refpar
%Strickland, D. K., \& Heckman, T. M. 2009, ApJ, 697, 2030
\refpar
Sturm, E., et al.\ 2011, ApJ,  733, L16 (S11)
% \refpar
% Surace, J., et al. 1998, ApJ, 492, 116
\refpar
Tacconi, L. J., et al. 2006, ApJ, 640, 228
\refpar
Tacconi, L. J., et al. 2010, Nature, 463, 781
\refpar
Teng, S. H., Veilleux, S., \& Baker, A. J. 2013, ApJ, 765, 95
% \refpar
% Teng, S. H., et al.\ 2008, ApJ, 674, 133
% \refpar
% Teng, S. H., et al.\ 2009, ApJ, 691, 261
% \refpar
% Teng, S. H., \& Veilleux, S. 2010, ApJ, 725, 1848
% \refpar
% Thompson, I., et al. 1980, MNRAS, 192, 53
% \refpar
% Trump, J. R. et al. 2006, ApJS, 165, 1.
% \refpar
% Turner, T. J. 1999, ApJ, 511, 142
% \refpar
% Turner, T. J., \& Kraemer, S. B. 2003, ApJ, 598, 916
% \refpar
% Ulvestad, J., Wrobel, J. M., \& Carilli, C. L. 1999, ApJ, 516, 127
\refpar 
Veilleux, S. 2012, J. Physics: Conference Series, 372, 012001
\refpar 
Veilleux, S., Cecil, G., \& Bland-Hawthorn, J.  2005, ARA\&A, 43, 769
\refpar
Veilleux, S., Kim, D.-C., \& Sanders, D. B. 2002, ApJS, 143, 315
\refpar
Veilleux, S., et al. 2006, ApJ, 643, 707
\refpar
Veilleux, S., et al.\ 2009a, ApJ, 701, 587 
\refpar
Veilleux, S., et al. 2009b, ApJS, 182, 628 (V09)
\refpar
Veilleux, S., et al. 2013, ApJ, 764, 15
\refpar
Weiner, B., et al. 2009, ApJ, 692, 187
\end{references}
\end{document}